\documentclass[epj]{svjour}
\usepackage{epsfig} 
\usepackage{times}
\usepackage[active]{srcltx}
\begin{document}

\title{\hfill {\footnotesize FZJ--IKP(TH)--2007--07} \\
The ${\bf{pp{\to}K^+\Sigma^+n}}$ cross section from missing mass spectra}

\author{A.~Sibirtsev\inst{1}, J.~Haidenbauer\inst{2}, H.-W.~Hammer\inst{1}
and Ulf-G.~Mei{\ss}ner\inst{1,2}}
\institute{Helmholtz-Institut f\"ur Strahlen- und Kernphysik
(Theorie), Universit\"at Bonn, Nu\ss allee 14-16, D-53115 Bonn,
Germany \and Institut f\"ur Kernphysik, Forschungszentrum J\"ulich,
D-52425 J\"ulich, Germany}

%\ead{a.sibirtsev@fz-juelich.de}

\abstract{
We utilize existing inclusive data on $K^+$-meson momentum spectra 
of the reaction $pp{\to}K^+X$ at $T_p$ = 2.3 - 2.85 GeV to deduce 
total cross sections for $pp{\to}K^+\Sigma^+n$.
The method used to extract those cross sections is explained and
discussed in detail. Our result for $T_p$ = 2.85 GeV is consistent 
with the data point from a direct measurement at the same beam energy.
The cross section obtained for $T_p$ = 2.3 GeV is with $13.7\pm2.3$ $\mu b$ 
considerably smaller than the value found in a recent experiment by
the COSY-11 Collaboration at a somewhat lower beam energy, indicating
that the $pp{\to}K^+\Sigma^+n$ reaction cross section could exhibit a
rather unusual energy dependence. 
}

\authorrunning{ A.~Sibirtsev {\it et. al} }
\titlerunning{The $pp{\to}K^+\Sigma^+n$ cross section from missing mass
spectra}

\maketitle

%%%%%%%%%%%%%%%%%%%%%%%%%%%%%%%%%%%%%%%%%%%%%%%%%%%%%%%%%%%%%%%%%%%%%%%%%%%%%%%%
\section{Introduction}
Recently first near-threshold cross section data for the $pp\to K^+\Sigma^+n$ 
reaction were published by the COSY-11 Collaboration~\cite{Rozek}. 
Surprisingly, it turned out that this cross section is larger than the
one for $pp{\to}K^+\Sigma^0p$ by a factor of around 230 at the
excess energy $\epsilon$=13~MeV and by a factor of around 90 at 
$\epsilon$=60~MeV. The excess energy $\epsilon$ is defined as
$\epsilon{=}\sqrt{s}{-}m_K-m_\Sigma -m_N$, where $s$ is the squared
invariant collision energy, while $m_K$, $m_\Sigma$ and $m_N$ are the
masses of the kaon, the $\Sigma$ hyperon and the nucleon, respectively. 
It is also worth mentioning that none of the available model calculations
\cite{Laget,Sibirtsev2,Tsushima,Shyam1,Zou} is able to describe those data.
In fact, most of those models underestimate the cross section by an order of 
magnitude or even more. 

Besides this rather large value for the production 
cross section as compared to the $\Sigma^0p$ channel, 
the new results for the $pp{\to}K^+\Sigma^+n$ reaction are also 
somewhat startling when compared with the available high energy data.
Indeed, one can find only five data 
points~\cite{Louttit,Sondhi,Dunwoodie,Chinowsky} 
for $pp{\to}K^+\Sigma^+n$ at higher energies in the literature.  
Moreover, those data show large fluctuations, 
even considering the large experimental uncertainties,
and two of those points~\cite{Dunwoodie} were reported only in a preprint. 
But it is still obvious that the COSY-11 result at $\epsilon$=60 MeV
\cite{Rozek}
is as large as the $pp{\to}K^+\Sigma^+n$ cross section measured at higher energies
\cite{Louttit,Sondhi,Chinowsky}, suggesting that there could be practically no 
energy dependence over the large energy region 60${\le}\epsilon{\le}1000$ MeV 
with the mean cross section being 49$\pm$5~$\mu$b. 
That is a rather unexpected result since the cross section 
of $pp{\to}K^+\Sigma^0p$, the only well investigated $\Sigma$ 
production channel, shows a significant energy dependence, as 
expected from the increasing phase space for the reaction. 
Indeed \, here the cross \, section changes by a factor of 
about 40 within the energy range indicated above. 

The data points~\cite{Louttit,Sondhi,Dunwoodie,Chinowsky} 
at high energies are obtained from bubble chamber images 
where the identification of the $pp\to K^+\Sigma^+n$ as well as 
the $pp{\to}K^+\Sigma^0p$ reaction channel was done simultaneously
and unambiguously. Therefore, these results at high energies
might be fairly reliable. 
The situation with regard to the more recent counter experiments is different. 
Here the $pp{\to}K^+\Sigma^+n$ channel was often not considered 
because of the substantial difficulties in the final particle
identification. The COSY-11 collaboration reconstructs the kaon and neutron
four-momenta and identifies the $\Sigma^+$-hyperon by the missing mass. 
It was found~\cite{Rozek} that the large background under the
$\Sigma^+$-signal complicates the data analysis considerably and it 
introduces large uncertainties. A much better, i.e. direct, identification 
of the $\Sigma^+$ can be done by detecting the $\Sigma^+{\to}p\pi^0$ decay
mode, though then a photon detector is required. Indeed, 
a corresponding experiment has been already proposed~\cite{Gillitzer}  
for the WASA detector~\cite{Adam} at the COSY facility.

With the present paper we want to supply some more values for the 
$pp{\to}K^+\Sigma^+n$ cross section to the data base. For that aim
we utilize available data on inclusive $K^+$-meson momentum spectra
measured at different angles in $pp$ collisions for the 
reaction $pp{\to}K^+ X$. Since 
the experimental $K^+$-meson momentum spectra~\cite{Hogan,Siebert,Reed}  
are available at energies that lie between the data of the COSY-11 
Collaboration and the high energy data, the result of our
analyis allows conclusions on the behavior of the $pp{\to}K^+\Sigma^+n$ 
cross section in this interesting energy region. 
Some of the spectra are available at energies that overlap with the 
bubble chamber results~\cite{Louttit} and, therefore, 
we can also check whether the results based on our method are compatible 
with the high energy measurement. As a byproduct we also provide 
cross sections for the $pp{\to}K^+\Lambda{p}$ reaction and compare 
them with direct measurements, where the latter are based on the reconstruction 
of the final particles.

The paper is organized as follows: In Sec. 2 we describe the
method. The analysis of the data is presented in Sec. 3. Our results are
compared to other available data in Sec. 4. The paper ends with
a short summary. 

%%%%%%%%%%%%%%%%%%%%%%%%%%%%%%%%%%%%%%%%%%%%%%%%%%%%%%%%%%%%%%%%%%%%%%%%%%%%%%%%
\section{Method for the data evaluation}
In this section we describe in detail the method for the data analysis. For 
completeness 
we include all relevant formulas, although some of them are given 
in Ref.~\cite{Byckling}. Furthermore, since this method is
not limited specifically to $pp{\to}K^+\Sigma^+n$ but applicable to any 
reaction with a three-body final state, we provide the formalism in a general 
form. The cross section for the $a{+}b{\to}1{+}2{+}3$
reaction is given by 
\begin{eqnarray}
\sigma{=}\frac{1}{2^6\pi^5\lambda^{1/2}(s,m_a^2,m_b^2)}
\!\int \!\frac{d^3p_1}{2E_1}\, \frac{d^3p_2}{2E_2}\, \frac{d^3p_3}{2E_3}\,
\nonumber  \\ \times
\delta(P_1{+}P_2{+}P_3{-}P_a{-}P_b)\, |{\cal A}|^2 \ ,
\end{eqnarray}
where $p_i$ and $E_i$ are the 3-momentum and the energy of the $i$-th
particle, respectively, while $P_i$ stands for the 4-momentum. 
${\cal A}$ denotes the reaction amplitude and 
the $\lambda$-function is defined by $\lambda(x,y,z)=(x-y-z)^2-4yz$.
We use the invariants
\begin{eqnarray}
s{=}P^2{=}(P_a+P_b)^2\!\!, \,\,\, \,\,\,  
s_Q{=}Q^2{=}(P_2+P_3)^2{=}(P-P_1)^2\!\!,
\end{eqnarray}
where $s_Q$ is the squared missing mass with respect to the first 
particle, which is identical to the squared invariant mass of the
second and third particle. The Lorentz invariant differential cross
section for the production of particle $1$ is then written as
\begin{eqnarray}
\frac{E_1}{p_1^2}\frac{d^3\sigma}{dp_1d\Omega_1}{=}
\frac{1}{2^7\pi^5\lambda^{1/2}(s,m_a^2,m_b^2)}
\int\!\!\frac{d^3p_2}{2E_2}\,\frac{d^3p_3}{2E_3}\nonumber \\
\times\delta(P_2+P_3-Q)\,\, |{\cal A}|^2 
= \frac{1}
{2^8\pi^4\lambda^{1/2}(s,m_a^2,m_b^2)} \nonumber \\
\times\frac{\lambda^{1/2}(s_Q,m_2^2,m_3^2)}
{s_Q}\,\, \overline{|{\cal A}|^2} \ . 
\label{spectrum1}
\end{eqnarray}
In the laboratory frame, i.e. for $P_b{=}({\vec
0},m_b)$,  $s_Q$ can be expressed as
\begin{eqnarray}
s_Q=s+m_1^2-2(E_a+m_b)E_1+2p_ap_1cos\theta_1,
\end{eqnarray}
where $\Omega_1$ and $\theta_1$ are the solid and polar production angle
of the first particle. In Eq.~(\ref{spectrum1}),
$\overline{|{\cal A}|^2}$ is the square of the  reaction amplitude integrated
over the kinematical variables related to the second and third
particle. In general $\overline{|{\cal A}|^2}$ depends on
$p_1$ (or $s_Q$), $\Omega_1{=}({\theta_1},{\phi_1})$ and $s$.

The relation between the differential momentum spectrum measured at
the solid angle $\Omega_1$ and the missing mass ($M_X$) spectrum, where
$M_X^2{=}s_Q$,  is given by
\begin{eqnarray}
\frac{d^3\sigma}{dM_X d\Omega_1}{=}\sqrt{s_Q}\, \left[\frac{E_a+m_b}
{\sqrt{p_1^2+m_1^2}}\,p_1{-}p_a\cos\theta_1\right]^{-1}\!\!\!
\frac{d^3\sigma}{dp_1d\Omega_1} \ .
\label{trans}
\end{eqnarray}

\begin{figure*}[t]
\vspace*{1mm}
\centerline{\hspace*{-1mm}\psfig{file=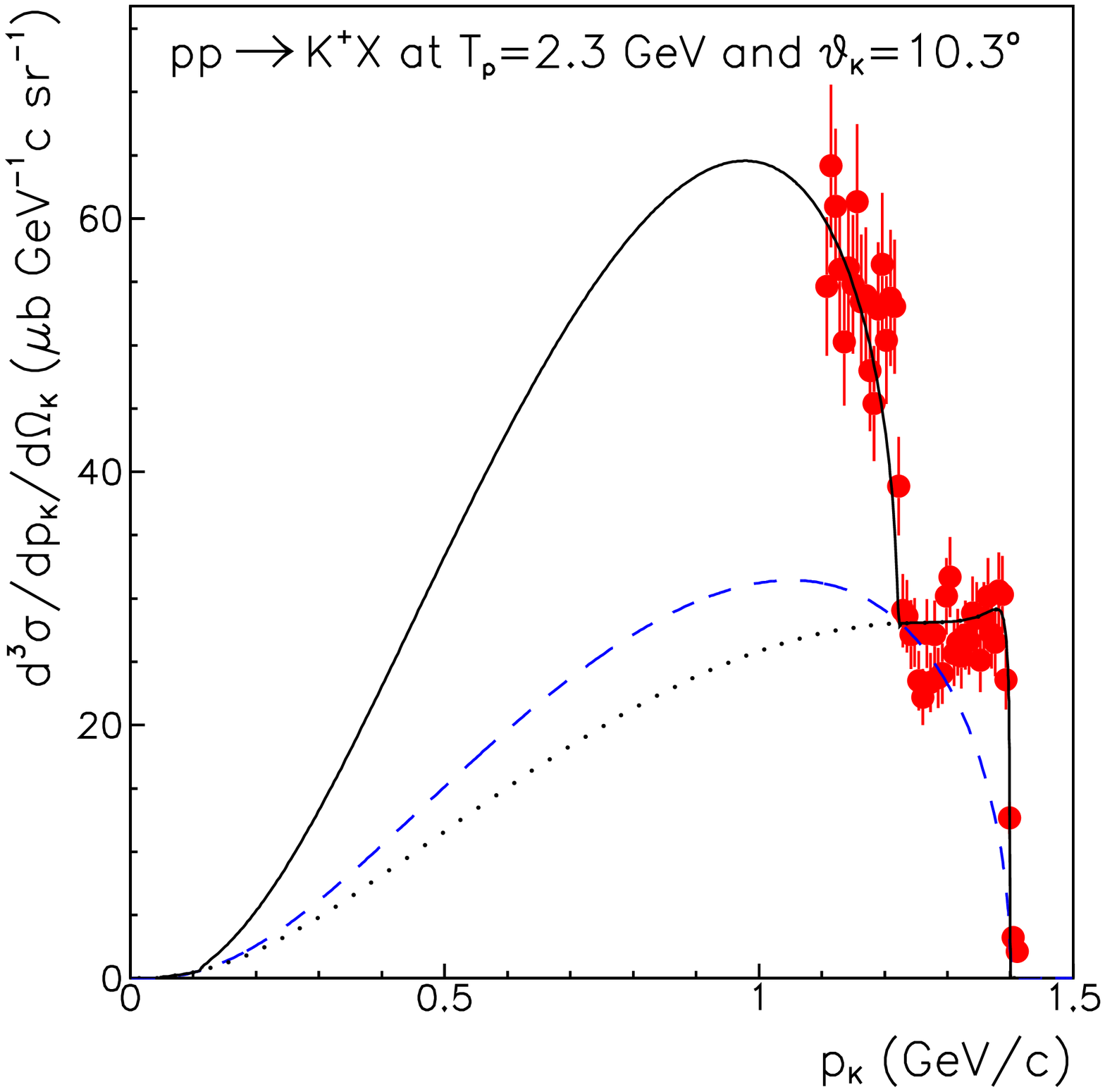,width=7.9cm,height=6.5cm}
\hspace*{-3mm}\psfig{file=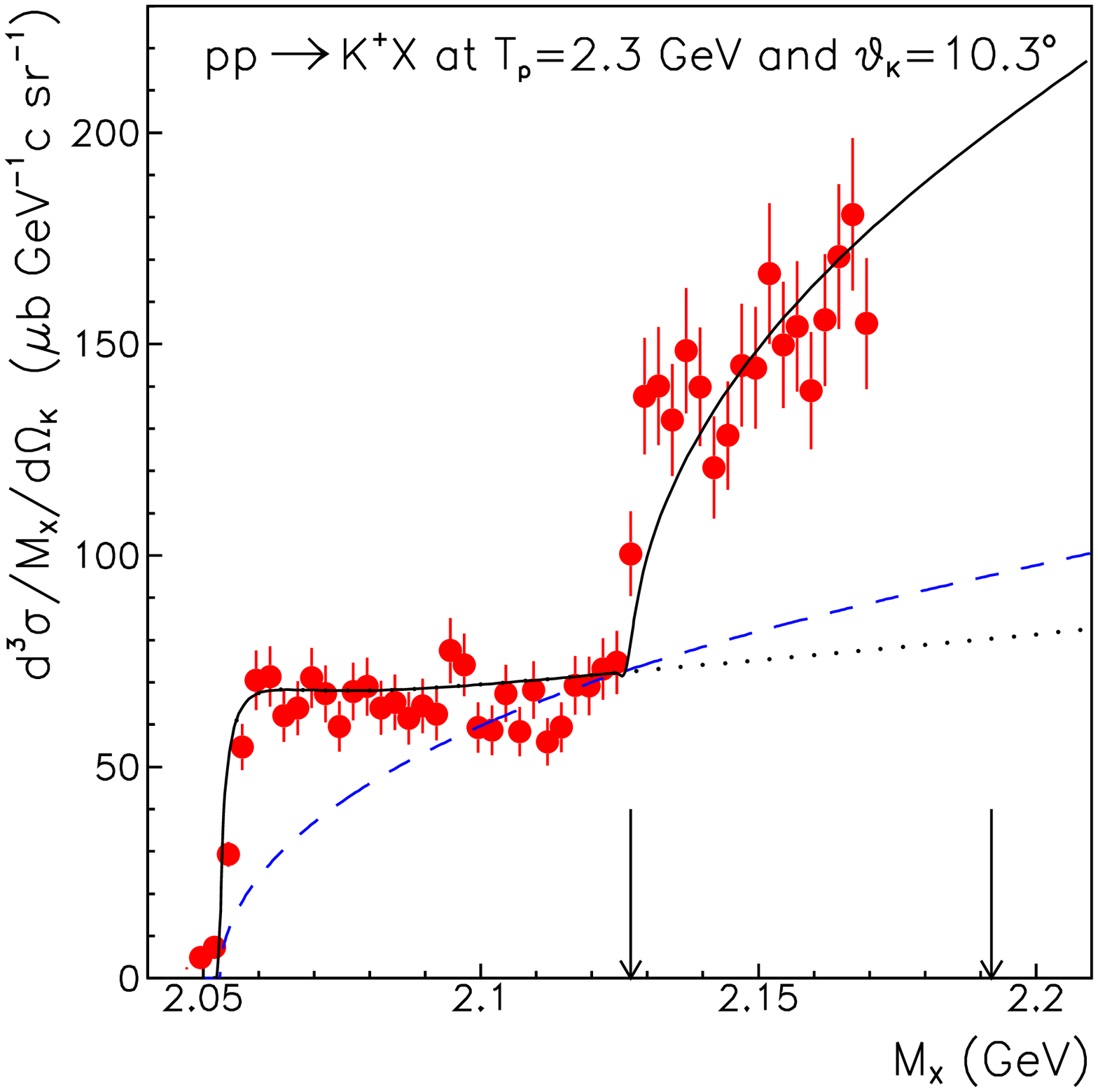,width=7.9cm,height=6.5cm}}\vspace*{-3mm}
\centerline{\hspace*{-1mm}\psfig{file=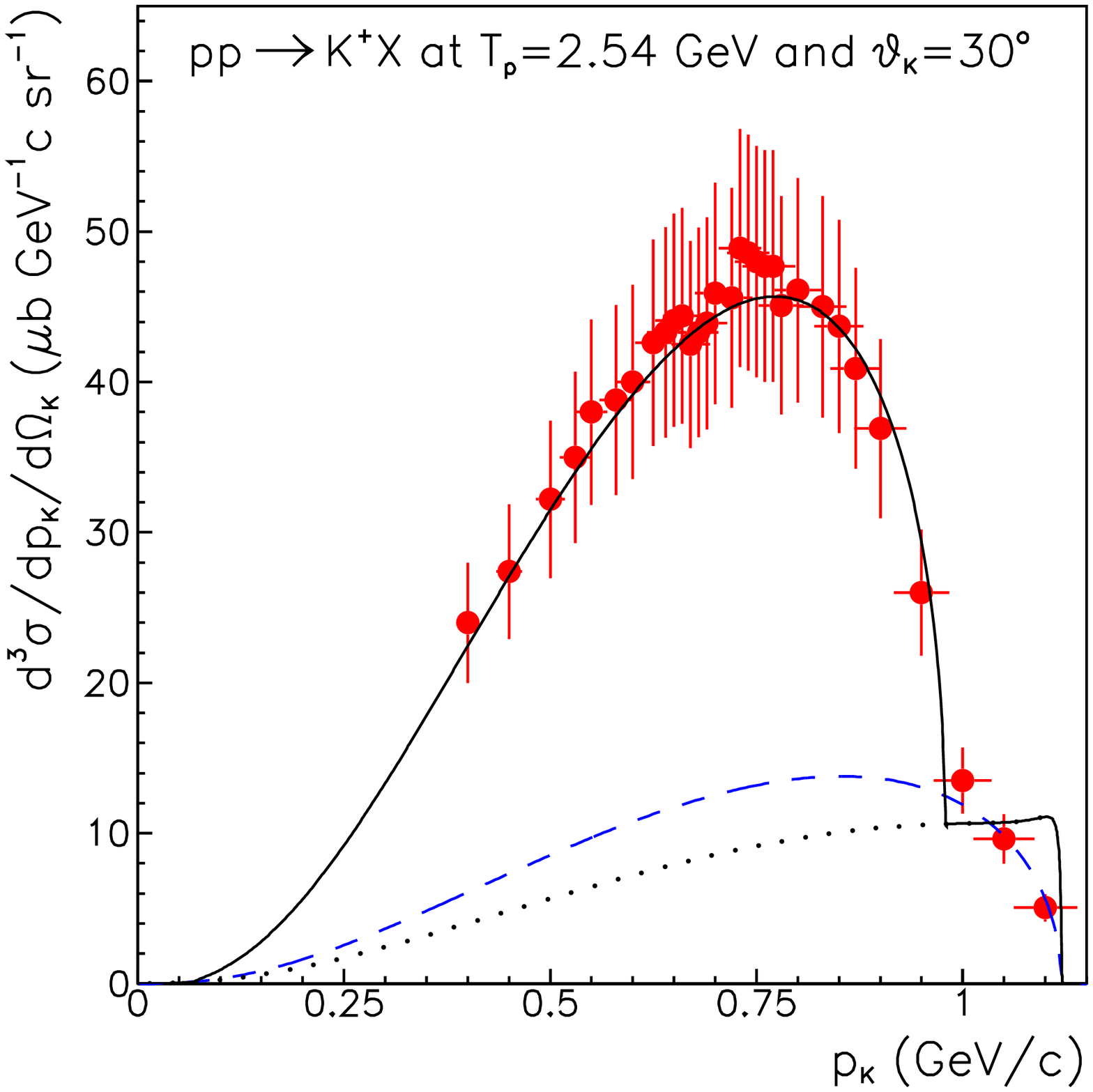,width=7.9cm,height=6.5cm}\hspace*
{ -3mm}\psfig{file=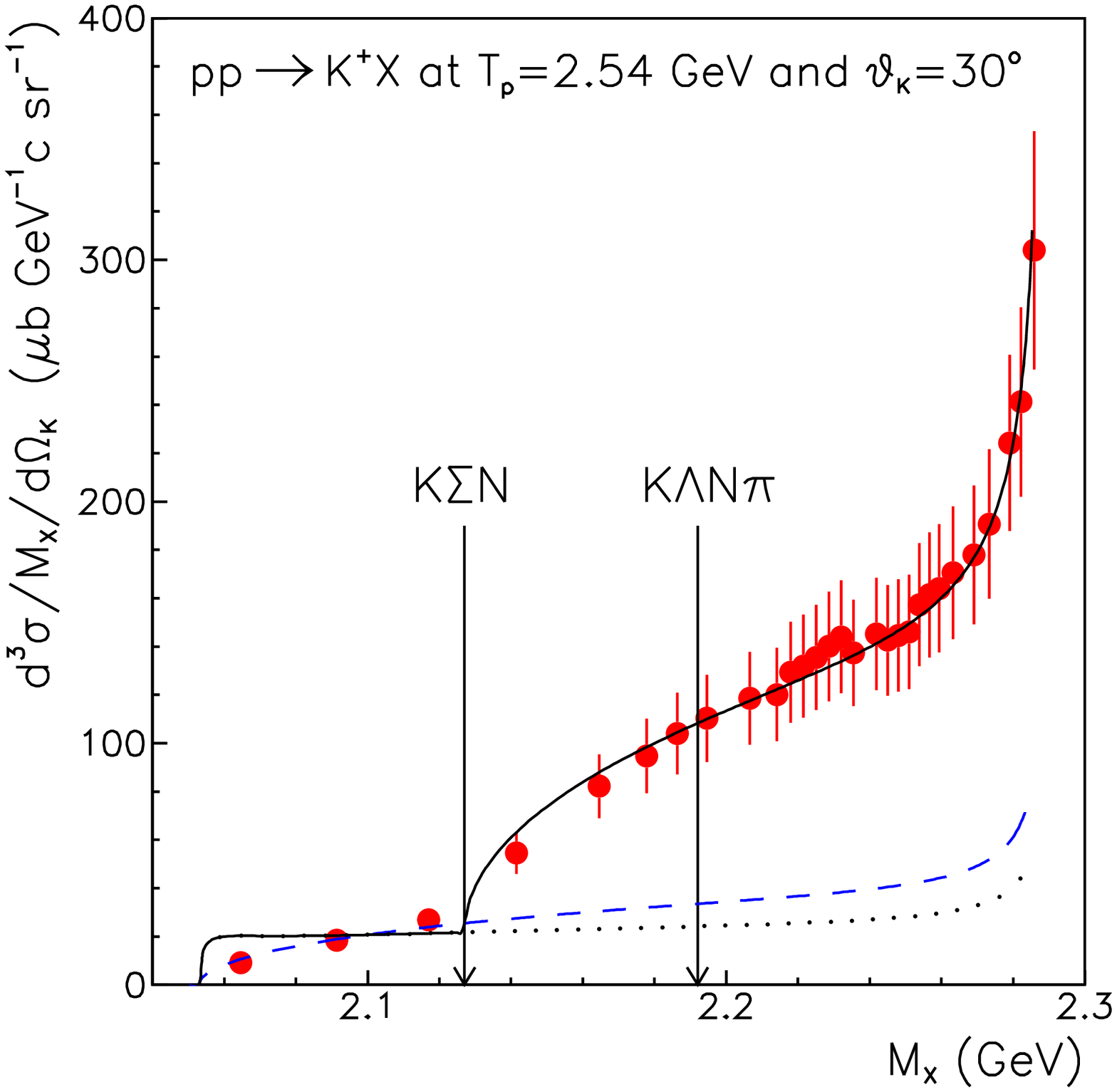,width=7.9cm,height=6.5cm}}
\vspace*{-3mm}
\caption{Left: Experimental information on the $K^+$-momentum spectra from the
$pp{\to}K^+X$ reaction at $T_p{=}$2.3 GeV and $\theta_K{=}10.3^o$~\cite{Siebert}
(upper) and at $T_p{=}$2.54 GeV and $\theta_K{=}30^o$~\cite{Hogan} (lower).
Right: Corresponding missing mass ($M_X$) spectra obtained by Eq.~(\ref{trans}). 
The arrows indicate the $K\Sigma{N}$ and $K\Lambda{N}\pi$ reaction
thresholds, respectively. 
The dotted lines show calculations based on Eqs.~(\ref{spectrum1}) and (\ref{trans}) 
for the $pp{\to}K^+\Lambda{p}$ reaction with $|{\cal A}_0|$ fitted to the data
for $M_X{\le}m_\Sigma{+}m_N$ and $\Lambda{p}$ FSI effects included via Eq.~(\ref{jost}).
The dashed lines are result obtained without inclusion of the $\Lambda{p}$ FSI. 
The solid lines are the sum of the $pp{\to}K^+\Lambda{p}$, $pp{\to}K^+\Sigma^0p$ and
$pp{\to}K^+\Sigma^+n$ cross sections where the sum of the cross sections for 
the two $\Sigma$-hyperon channels was fitted to the data for 
$m_\Sigma{+}m_N \le{M_X}{\le}m_\Lambda{+}m_N{+}m_\pi$.
} 
\label{kapro6e_bm}
\end{figure*}

In order to exemplify how we proceed in our analysis let us consider here two 
typical data samples for the $pp{\to}K^+X$ reaction. One is from a measurement at 
the beam energy $T_p$=2.3 GeV and the $K^+$-production angle $\theta_K{=}10.3^o$
\cite{Siebert} and the other at $T_p$=2.54~GeV and $\theta_K{=}30^o$
\cite{Hogan}. Both data sets are shown in Fig.~\ref{kapro6e_bm}, where
the left panel illustrates the $K^+$-meson momentum spectra, 
while the right panel shows the missing mass spectra obtained by 
Eq.~(\ref{trans}). The arrows in 
Fig.~\ref{kapro6e_bm} indicate the $K\Sigma{N}$ and $K\Lambda{N}\pi$ reaction
thresholds, respectively. Below the $K\Sigma{N}$ threshold, i.e. for 
$M_X{\le}m_\Sigma{+}m_N$, kaon production is primarily due to the
$pp{\to}K^+\Lambda{p}$ reaction, though contributions from that
channel with additional photons ($K^+\Lambda{p}\gamma$, etc.) are 
also possible. 
The contributions to the missing mass spectrum for 
$m_\Sigma{+}m_N \le{M_X}{\le}m_\Lambda{+}m_N{+}m_\pi$ come from the
$pp{\to}K^+\Lambda{p}$, $pp{\to}K^+\Sigma^0p$ and $pp{\to}K^+\Sigma^+n$
reaction channels and again channels with additional 
photons. Thus, by subtracting the contribution of $pp{\to}K^+\Lambda{p}$ 
in that invariant-mass region one can extract the sum of the
$pp{\to}K^+\Sigma^0p$ and $pp\to K^+\Sigma^+n$ channels, under the
assumption that the reactions with photons provide a negligible contribution. 
For a reliable estimation of the $pp{\to}K^+\Lambda{p}$ contribution 
it is crucial to know the $K^+$-meson spectra below the
$K\Sigma{N}$ threshold. Then one can fix the $pp{\to}K^+\Lambda{p}$ channel 
directly from those data and use that result for the extrapolation to
the invariant-mass region where the $\Sigma$ channels are open. 
As is clear from Fig.~\ref{kapro6e_bm} in some of the available experiments 
there are only a few data points below the $K\Sigma{N}$ threshold. In 
such a case there is a sizable uncertainty in the data evaluation, which will
be reflected, eventually, in the error bars of the corresponding results.
This uncertainty is somewhat reduced if there is at least a clean signal 
of the opening of the $K\Sigma N$ threshold in the spectrum, which is the 
case for most of the data. 

The $pp{\to}K^+\Sigma^0p$ reaction can be well identified by the
detection of the final particles and it has been intensively
investigated \cite{Sewerin1,Kowina1}. This information can then be used 
to deduce the contribution of the $pp{\to}K^+\Sigma^+n$ channel from the 
missing mass spectra of inclusive $K^+$-meson production in the region 
$m_\Sigma{+}m_N \le{M_X}{\le}m_\Lambda{+}m_N{+}m_\pi$ 
for various beam energies. As is obvious from 
Fig.~\ref{kapro6e_bm}, while some experiments provide information on 
the $K^+$-meson momentum spectrum over basically the whole available 
phase space, this is not the case with others. These limitations again 
introduce an uncertainty in the data analysis.

For fixing the contribution of the $pp{\to}K^+\Lambda{p}$ reaction 
channel to the missing mass spectrum we utilize Eqs.~(\ref{spectrum1}) 
and (\ref{trans}). Specifically, we determine the reaction amplitude 
$|{\cal A}|$ from the data for $M_X{\le}m_\Sigma{+}m_N$ and use that value
for the extrapolation to $M_X{\ge}m_\Sigma{+}m_N$. 
Since it is well known that the $\Lambda{p}$ final-state interaction (FSI) 
is sizable \cite{Sibirtsev1,Sibirtsev1a} we take it into account explicitly.
This is done by assuming \cite{Watson,Migdal} that the reaction amplitude 
${\cal A}$ can be 
factorized into a practically momentum and energy independent elementary 
production amplitude ${\cal A}_0$ and an FSI factor, ${\cal A}_{\Lambda p}$, 
where the latter is calculated within the Jost-function approach
\cite{Goldberger}. Details are summarized in Appendix \ref{App}. 
As mentioned there, in our analysis we use Jost-function parameters (or 
equivalently, effective range parameters) from a global fit to 
the reaction $pp{\to}K^+\Lambda{p}$. 
In principle, one could try to determine those parameters for each data
set separately by using the corresponding $M_X$-spectra 
\cite{Hinterberger,Gasparyan1,Gasparyan2}.
But then 
one would require $M_X$-spectra with rather good mass resolution and
statistical accuracy, which is not always the case as seen in
Fig.~\ref{kapro6e_bm} for $T_p$=2.54~GeV. Consequently, only some of the 
data could be analyzed by including FSI effects that are determined 
by fitting directly to those data themselves. 

Thus, we only fix the elementary production amplitude ${\cal A}_0$
by a fit to the corresponding data for $M_X{\le}m_\Sigma{+}m_N$. In 
practice we determine the constant $|{\cal A}_0|$ for each angle and beam
energy where experimental $K^+$ mass spectra are available. Then, by
averaging the obtained values for $|{\cal A}_0|$ (at a specific energy) over 
the angular dependence we deduce a result for $\overline{|{\cal A}_0|^2}$.
The latter quantity can then be compared with the amplitudes deduced from 
directly measured $pp{\to}K^+\Lambda{p}$ cross sections, cf. the discussion 
in \ref{App} and the results presented in Sect. 4. This allows us 
to examine whether the results we extracted from the measured invariant-mass 
spectra are consistent with the experimental information
on the total $\Lambda$ production cross sections. 

Let us now come back to the $pp{\to}K^+X$ invariant mass spectrum. 
The dotted lines in Fig.~\ref{kapro6e_bm} show results of a calculation 
based on Eqs.~(\ref{spectrum1}) and (\ref{trans}) with the FSI included
via Eq.~(\ref{jost}) and the squared reaction amplitude 
$|{\cal A}_0|$ appropriately adjusted to the spectra at $M_X{<}m_\Sigma{+}m_N$. 
The description of the $K^+$-meson momentum and missing mass spectra in terms 
of the contribution from the $pp{\to}K^+\Lambda{p}$ reaction looks reasonable. 
Note that so far we have neglected possible contributions from the reactions 
with photons in the final state, i.e. $pp{\to}K^+\Lambda{p}\gamma$, 
$pp{\to}K^+\Lambda{p}\gamma\gamma$ etc. However, judging from the measurement
where a decent number of data points is available for $M_X < m_\Sigma{+}m_N$,
there is not much room for such additional contributions anyway.

In order to estimate the uncertainty 
that could arise from our treatment of the $\Lambda p$ 
FSI we consider also an alternative procedure. We perform a fit to the 
$K^+$ invariant mass spectrum without FSI, i.e. with pure phase space. 
But in this case we consider only data points that lie in an energy interval
of about 30 MeV from the $K\Sigma N$ threshold downwards
for the determination of the reaction amplitude $|{\cal A}_0|$ at
the various angles and energies.
The data points closer to the $K\Lambda N$ threshold exhibit, in general, such 
obvious FSI effects that it is meaningless to try to fit them with pure phase space. 
The dashed lines in Fig.~\ref{kapro6e_bm} show those results obtained without 
inclusion of the $\Lambda{p}$ FSI. 
We will use the predictions of those fits for the $K\Lambda N$ invariant
mass spectrum in the region $m_\Sigma{+}m_N{\le}M_X{\le}m_\Lambda{+}m_N{+}m_\pi$
for extracting the $\Sigma$ production cross section too. 
However, we want to emphasize already at this stage that we consider the 
extrapolation based on the fit that includes the $\Lambda{p}$ FSI as much more
reliable and, therefore, we consider the $\Sigma$ cross sections deduced
from that fit as our definitive results. 

Once the contribution from the $pp{\to}K^+\Lambda{p}$ channel is 
established we subtract it from the $pp{\to}K^+X$ data 
in the region $m_\Sigma{+}m_N{\le}M_X{\le}m_\Lambda{+}m_N{+}m_\pi$ 
in order to obtain the sum of the contributions from the
$pp{\to}K^+\Sigma^0{p}$ and $pp{\to}K^+\Sigma^+{n}$
reactions. Possible additional contribution from channels with
photons in the final state are again neglected. 
Utilizing again Eqs.~(\ref{spectrum1}) and (\ref{trans}) we  
determine the corresponding (combined) $\Sigma$ amplitude $|{\cal A}_0|$ 
for each angle and total energy where experimental $K^+$ mass spectra 
are available, etc. 
However, unlike in the $K^+\Lambda{p}$ channel, now we do not include
an FSI factor in the fitting procedure. Indeed, none of the available 
data sets exhibits a pronouncend enhancement near the $K\Sigma N$ 
threshold that would warrant the inclusion of FSI effects. 
The solid lines in Fig.~\ref{kapro6e_bm} show the final result, i.e. 
the contribution from the $pp{\to}K^+\Lambda{p}$ reaction plus the
fitted contribution from the $pp{\to}K^+\Sigma N$ channels. 

%%%%%%%%%%%%%%%%%%%%%%%%%%%%%%%%%%%%%%%%%%%%%%%%%%%%%%%%%%%%%%%%%%%%%%%%%%%%%%%%
\section{Data analysis}

In the present paper we analyze the measured $K^+$-meson momentum 
spectra published in Refs.~\cite{Hogan,Siebert,Reed}. 
The achieved results are summarized in Tables \ref{TAB1a}
and \ref{TAB1}. In order to stay as close as possible to physical
quantities we do not list the values obtained for the amplitudes
$|{\cal A}_0|$ but the corresponding cross sections. However, since
those amplitudes correspond to data at different angles it is obvious 
that the given values are not really total cross sections. 
Rather, they represent cross sections for specific angles, appropriately 
normalized to the full solid angle. In order to remind the reader on that 
we put the superscript $\theta$ on the corresponding symbols 
($\sigma_\Lambda^\theta$ or $\sigma_\Sigma^\theta$). 
\begin{table*}
\caption{\label{TAB1a}
Analysis of available data on $K^+$-meson inclusive momentum spectra 
from the $pp{\to}K^+X$ reaction: Results for $pp{\to}K^+\Lambda{p}$.
Specified are the proton beam energy $T_p$, the kaon production 
angle $\theta_K$, and the excess energy $\epsilon$ with respect to the 
$\Lambda$-hyperon production threshold.
$\sigma_\Lambda^\theta$ is the $pp{\to}K^+\Lambda{p}$ reaction cross
section obtained from a fit to the data at a specific angle $\theta_K$, as
explained in the text. The factor $\xi$ indicates the angle dependence, 
cf. text. 
}
\footnotesize\rm
\begin{tabular*}{\textwidth}{@{}l*{15}{@{\extracolsep{0pt plus12pt}}l}}
\hline\noalign{\smallskip}
Reference & $T_p$ (GeV) & $\epsilon$ (MeV) & $\theta_K$ (degrees) 
 & w/o $\Lambda{p}$ FSI& \multicolumn{2}{l}{with $\Lambda{p}$ FSI}\\
 & & &  & 
 $\sigma_\Lambda^\theta$ ($\mu$b) & $\xi$ & $\sigma_\Lambda^\theta$ ($\mu$b)\\
\noalign{\smallskip}\hline\noalign{\smallskip}
\cite{Siebert} & 2.3 & 252 & 8.3 & 
23.9$\pm$0.7 & 0.80 & 24.2$\pm$0.9 \\ 
\cite{Siebert} & 2.3 & 252 & 10.3 &21.3$\pm$0.6 & 0.72 &
21.8$\pm$0.8 \\ 
\cite{Siebert} & 2.3 & 252 & 12.0  &20.0$\pm$0.6 & 0.68 & 
20.6$\pm$0.6\\ 
\noalign{\smallskip}\hline\noalign{\smallskip}
\cite{Reed} & 2.4 & 285 & 0  & 77.0$\pm$8.9  & 2.1 & 
72.9$\pm$9.2 \\ 
\cite{Reed} & 2.4 & 285 & 17 
& 52.8$\pm$6.7 & 0.9 & 31.3$\pm$8.1 \\ 
\noalign{\smallskip}\hline\noalign{\smallskip}
\cite{Hogan} & 2.54 & 331 & 20 & 
31.1$\pm$3.7 & 0.55 & 22.3$\pm$2.9\\ 
\cite{Hogan} & 2.54 & 331 & 30 & 
20.5$\pm$2.1 & 0.40 & 16.2$\pm$2.3  \\ 
\cite{Hogan} & 2.54 & 331 & 40  & 
24.2$\pm$2.3 & 0.40 & 16.2$\pm$2.7\\
\noalign{\smallskip}\hline\noalign{\smallskip}
\cite{Siebert} & 2.7 & 383 & 12.6 & 
32.7$\pm$0.3 & 0.6 & 28.0$\pm$0.6 \\ 
\cite{Siebert} & 2.7 & 383 & 16.1  & 
30.1$\pm$0.7  & 0.55 & 25.7$\pm$0.5\\ 
\cite{Siebert} & 2.7 & 383 & 20  & 
30.3$\pm$0.4 & 0.43 & 20.1$\pm$0.5\\ 
\cite{Siebert} & 2.7 & 383 & 23.5  & 
20.9$\pm$0.5  & 0.4 & 18.7$\pm$0.6\\
\noalign{\smallskip}\hline\noalign{\smallskip}
\cite{Reed} & 2.85 & 431 & 0 &  120.3$\pm$10.8 & 1.7 & 
88.0$\pm$11.2 \\ 
\cite{Reed} & 2.85 & 431 & 17 
& 39.6$\pm$4.8 & 0.5 & 25.9$\pm$4.2 \\ 
\cite{Reed} & 2.85 & 431  & 32 
& 28.6$\pm$6.1 & 0.5 & 25.9$\pm$5.3\\
\hline\noalign{\smallskip}
\end{tabular*}
\end{table*}

Comparing the resulting values for $\sigma_\Lambda^\theta$ and 
$\sigma_\Sigma^\theta$ at different angles (at a specific energy) 
allows conclusions on the angular dependence of the reaction. 
For facilitating an easy general examination of that dependence we 
introduce the quantity $\xi$ which is the ratio of $\sigma_\Lambda^\theta$
and the corresponding (genuine) total cross section $\sigma_\Lambda$ obtained 
from the reference amplitude (\ref{par2}) in conjuction with Eq.~(\ref{eval1}).
Evidently, if there is full consistency between the latter parametrization of 
the experimental total cross section and the result stemming from our 
evaluation of the missing mass spectrum then the average of $\xi$ over the 
kaon angles would amount to unity. 

Note that we have neglected the difference between the $p$ and $n$ masses and 
between the $\Sigma^0$ and $\Sigma^+$ masses in calculating the excess
energies. 
These are inessential at the high reaction energies we are dealing with here. 

\begin{table*}
\caption{\label{TAB1}
Analysis of available data on $K^+$-meson inclusive momentum spectra 
from the $pp{\to}K^+X$ reaction: Results for $pp{\to}K^+\Sigma N$.
Specified are the proton beam energy $T_p$, the kaon production 
angle $\theta_K$, and the excess energy $\epsilon$ with respect to the 
$\Sigma$-hyperon production threshold.
$\sigma_\Sigma^\theta$ is the sum of the $pp{\to}K^+\Sigma^0p$ and 
$pp{\to}K^+\Sigma^+n$ reaction cross sections obtained from 
a fit to the data at a specific angle $\theta_K$, as explained in 
the text. 
}
\footnotesize\rm
\begin{tabular*}{\textwidth}{@{}l*{15}{@{\extracolsep{0pt plus12pt}}l}}
\hline\noalign{\smallskip}
Reference & $T_p$ (GeV) & $\epsilon$ (MeV) & $\theta_K$ (degrees) 
& \multicolumn{2}{c}{w/o $\Lambda{p}$ FSI}
& \multicolumn{2}{c}{ with $\Lambda{p}$ FSI}\\
 & & & & $\sigma_\Sigma^\theta$ ($\mu$b) & 
$\chi^2$/ndf & $\sigma_\Sigma^\theta$ ($\mu$b) & $\chi^2$/ndf\\
\noalign{\smallskip}\hline\noalign{\smallskip}
\cite{Siebert} & 2.3 & 178 & 8.3  & 
13.7$\pm$1.0 & 1.3&  14.7$\pm$0.9 & 0.9\\ 
\cite{Siebert} & 2.3 & 178 & 10.3 &16.9$\pm$1.1 & 1.5& 
18.0$\pm$1.0 & 1.6 \\ 
\cite{Siebert} & 2.3 & 178 & 12.0  &20.0$\pm$1.2 & 1.2& 
19.5$\pm$1.0 & 2.1\\
\noalign{\smallskip}\hline\noalign{\smallskip}
\cite{Reed} & 2.4 & 211 & 0  & 32.5$\pm$12.9 & 0.5 & 
36.1$\pm$10.8 & 0.1\\ 
\cite{Reed} & 2.4 & 211 & 17 
& 19.0$\pm$6.1 & 0.2 &  50.6$\pm$13.2 & 1.2\\
\noalign{\smallskip}\hline\noalign{\smallskip}
\cite{Hogan} & 2.54 & 257 & 20 & 
32.0$\pm$2.4 & 0.6 & 41.4$\pm$2.5 & 0.1\\ 
\cite{Hogan} & 2.54 & 257 & 30 & 
41.0$\pm$1.8 & 0.1 &  45.7$\pm$1.8 & 0.2\\ 
\cite{Hogan} & 2.54 & 257 & 40 & 
34.2$\pm$2.8 & 3.5 &  42.4$\pm$2.8 & 0.4\\
\noalign{\smallskip}\hline\noalign{\smallskip}
\cite{Siebert} & 2.7 & 309 & 12.6  & 
37.9$\pm$0.8 & 5.7  &  47.8$\pm$0.5 & 7.6\\ 
\cite{Siebert} & 2.7 & 309 & 16.1  & 
50.4$\pm$1.4 & 2.9 &  54.2$\pm$1.4 & 2.6\\ 
\cite{Siebert} & 2.7 & 309 & 20 & 
51.1$\pm$0.9 & 5.5 &  62.7$\pm$0.9 & 4.2\\ 
\cite{Siebert} & 2.7 & 309 & 23.5 & 
28.4$\pm$1.5 & 1.7 &  30.6$\pm$1.1 & 1.6\\
\noalign{\smallskip}\hline\noalign{\smallskip}
\cite{Reed} & 2.85 & 357 & 0 & 60.1$\pm$6.8 & 1.3 & 
82.5$\pm$14.3 & 0.1 \\ 
\cite{Reed} & 2.85 & 357 & 17 
& 76.7$\pm$10.0 & 0.5 &  73.4$\pm$14.9 & 0.3\\ 
\cite{Reed} & 2.85 & 357 & 32 
& 23.4$\pm$7.8 & 2.0 & 20.0$\pm$8.6 & 0.8\\
\hline\noalign{\smallskip}
\end{tabular*}
\end{table*}

As already mentioned, a major source for systematical uncertainties 
in the data analysis by the method described above is due to the 
extrapolation of the $K^+\Lambda{p}$ mass spectrum to the region
$m_\Sigma{+}m_N{\le}M_X{\le}m_\Lambda{+}m_N{+}m_\pi$. 
It is clear from Fig.~\ref{kapro6e_bm} that the parametrization 
including the $\Lambda p$ FSI still differs from the pure phase-space
behavior in that region and, therefore, it affects the absolute value of 
the extracted $\Sigma$ production cross section. 
Thus, in order to estimate the uncertainty due to the extrapolation 
we determine $\sigma_\Sigma^\theta$ (by a fit based on 
Eqs.~(\ref{spectrum1}) and (\ref{trans})) for two scenarios: We
subtract the contribution from the $pp{\to}K^+\Lambda{p}$ (i) including 
FSI effects as shown by the dotted line in Fig.~\ref{kapro6e_bm}, 
and (ii) without FSI as given by the dashed line in Fig.~\ref{kapro6e_bm}. 
Both results for $\sigma_\Sigma^\theta$ are given in Table~\ref{TAB1}. 

We should mention that even 
within the Jost-function approach it is not always possible to reproduce 
the $M_X$-spectra around the $K^+\Lambda{p}$ threshold in a perfect way. 
That might be a problem related to the use of the Jost function formalism or 
simply due to uncertainties of the parameters $\alpha$ and $\beta$ (\ref{paral})
used in Eq.~(\ref{jost}). In any case, such more subtle aspects of the 
$\Lambda{p}$ FSI do not influence the shape of the $M_X$-spectra above the 
$K^+\Sigma{N}$ threshold significantly and are, therefore, not relevant
for us. 

\begin{figure*}[t]
\vspace*{1mm}
\centerline{\hspace*{-3mm}\psfig{file=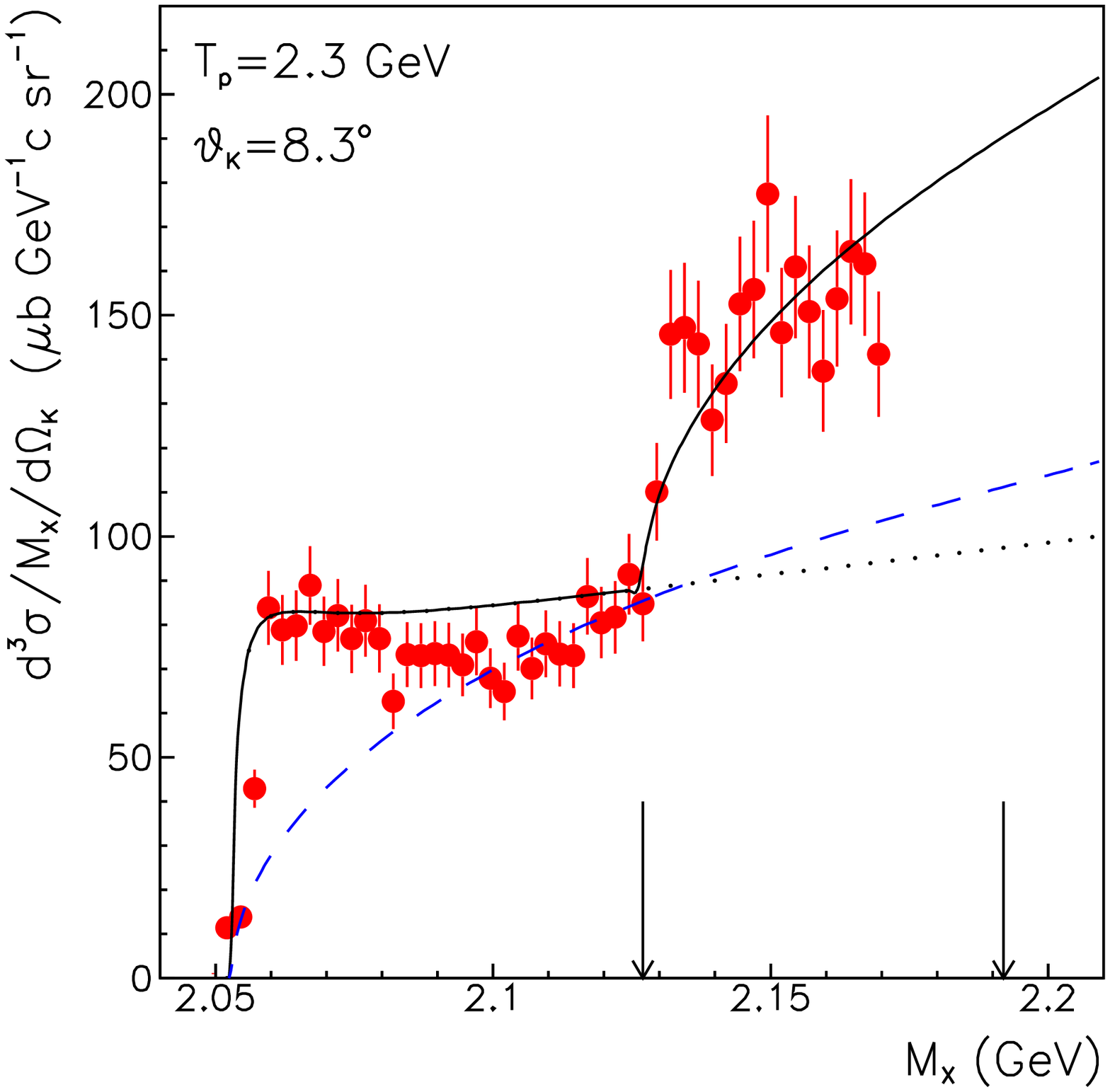,width=5.5cm,height=7.cm}
\hspace*{-5mm}\psfig{file=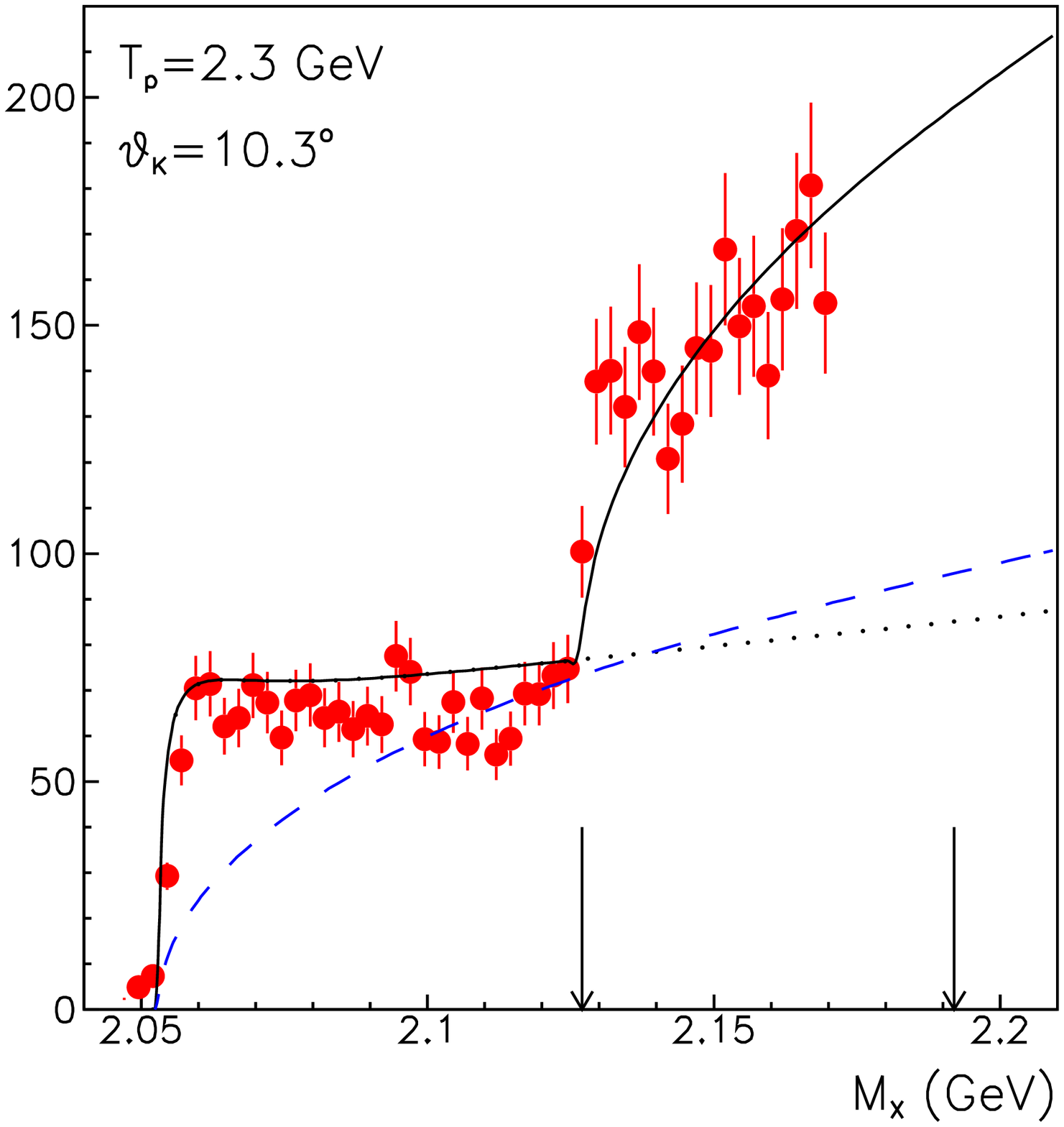,width=5.5cm,height=7.cm}\hspace*{-5mm}
\psfig{file=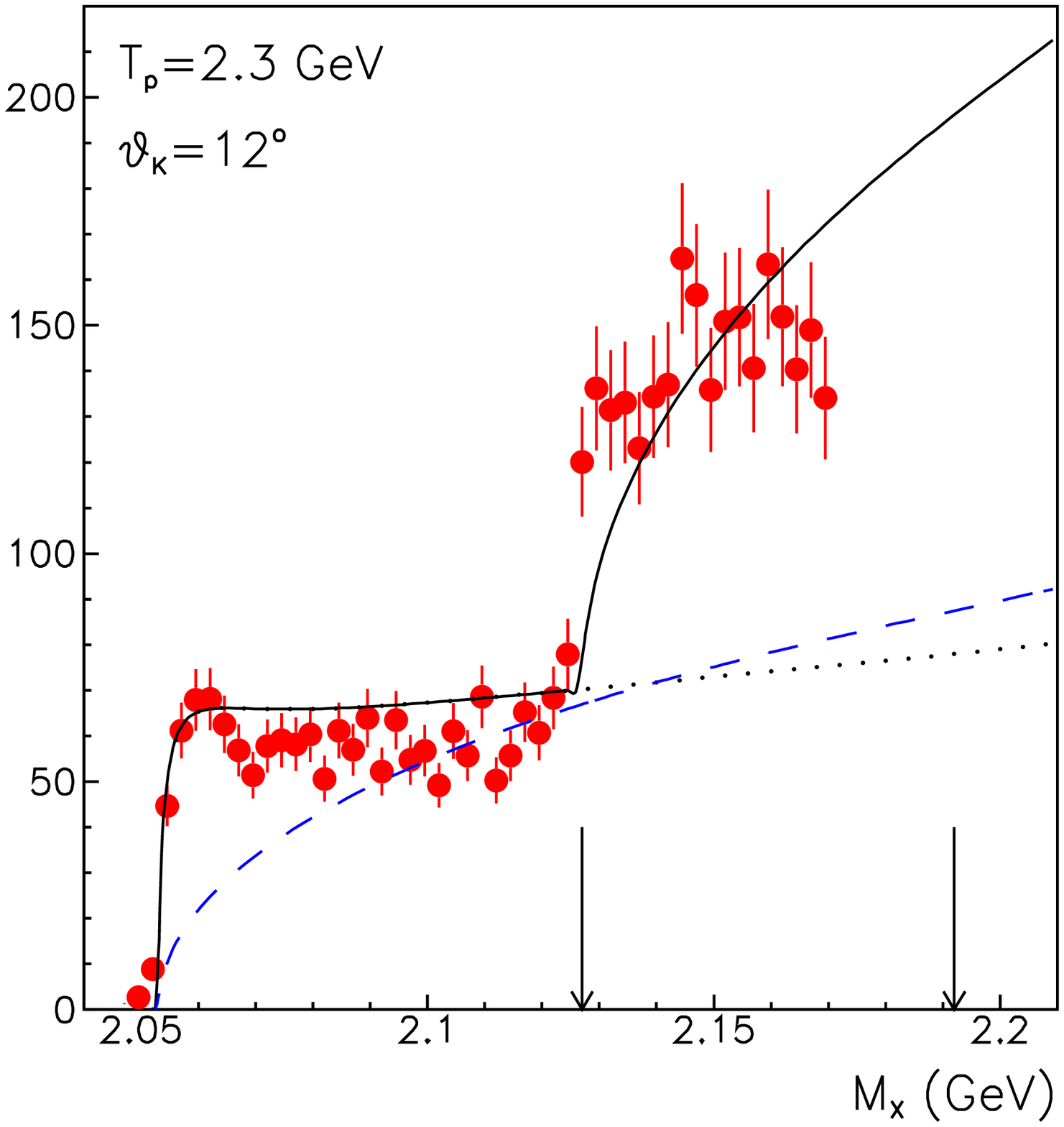,width=5.5cm,height=7.cm}}
\vspace*{-2mm}
\caption{Experimental missing mass
spectra obtained by Eq.~(\ref{trans}) from the $K^+$-meson momentum 
spectra of the $pp{\to}K^+X$ reaction~\cite{Siebert} at the proton beam energy 
of $T_p{=}$2.3 GeV at different kaon production angles $\theta_K$.
The arrows indicate the $K\Sigma{N}$ and $K\Lambda{N}\pi$ reaction
thresholds, respectively. 
The dotted lines show calculations based on Eqs.~(\ref{spectrum1}) and (\ref{trans}) 
for the $pp{\to}K^+\Lambda{p}$ reaction with $|{\cal A}_0|$ fitted to the data
for $M_X{\le}m_\Sigma{+}m_N$ and $\Lambda{p}$ FSI effects included via Eq.~(\ref{jost}).
The dashed lines are result obtained without inclusion of the $\Lambda{p}$ FSI. 
The solid lines are the sum of the $pp{\to}K^+\Lambda{p}$, $pp{\to}K^+\Sigma^0p$ and
$pp{\to}K^+\Sigma^+n$ cross sections where the sum of the cross sections for 
the two $\Sigma$-hyperon channels was fitted to the data for 
$m_\Sigma{+}m_N \le{M_X}{\le}m_\Lambda{+}m_N{+}m_\pi$.
} 
\label{kapro6e}
\end{figure*}

Let us now discuss the different data sets for the $pp{\to}K^+X$ reaction 
one by one. Fig.~\ref{kapro6e} shows the missing mass spectra from 
Ref.~\cite{Siebert} at the proton beam energy of
$T_p{=}$2.3 GeV and kaon production angles of $\theta_K{=}8.3^o$,
10.3$^o$ and 12$^o$. The dashed lines are the results for the 
$pp{\to}K^+\Lambda{p}$ reaction fitted with a constant reaction amplitude 
$|{\cal A}_0|$ alone, while the dotted lines indicate corresponding 
results including the $\Lambda{p}$ FSI.

The $M_X$-spectra exhibit a substantial enhancement close to the
$K^+\Lambda{p}$ threshold that originates from the $\Lambda{p}$ FSI. 
Note that the shape of the near threshold spectra depends somewhat on the
$K^+$-meson production angle and is not reproduced perfectly by using 
the Jost function (Eq.~(\ref{jost})), especially at the angles $\theta{=}8.3^o$
and 12$^o$. As was shown in Ref.~\cite{Hinterberger} the $M_X$-dependence 
generated by Eq.~(\ref{jost}) can be varied by changing the
parameters $\alpha$ and $\beta$ and, in principle, it is possible to 
achieve a better description of the spectra around the $K^+\Lambda{p}$
threshold by allowing $\alpha$ and $\beta$ to depend on the $\theta_K$ 
angle. However, all those variations have only a marginal influence on the 
description of the missing mass spectra above the $K^+\Sigma{N}$ threshold, 
which is the region we are interested in in the present analysis. 

The solid lines in Fig.~\ref{kapro6e} show the sum of the $pp{\to}K^+\Lambda{p}$, 
$pp{\to}K^+\Sigma^0p$ and $pp{\to}K^+\Sigma^+n$ channels where the 
contribution of the latter two channels was determined by a fit to the 
difference between the experimental spectra and the contribution 
from the $pp{\to}K^+\Lambda{p}$ reaction (including FSI effects) 
via Eqs.~(\ref{spectrum1}) and (\ref{trans}). 
In Tables~\ref{TAB1a} and \ref{TAB1} we list the corresponding values for 
$\sigma_\Lambda^\theta$ and $\sigma_\Sigma^\theta$ determined from the 
$M_X$-spectra for the cases with and without inclusion of the $\Lambda{p}$ FSI. 
The quality of the least-square fit can be judged from the given reduced $\chi^2$. 
As can be seen from the table, there is practically no difference between
the results obtained with and without $\Lambda{p}$ FSI. This is primarily due to
the fact that there are sufficient and accurate data on the mass spectrum below 
the $K\Sigma N$ threshold. Moreover, this threshold is clearly mapped out.

\begin{figure*}[t]
\vspace*{1mm}
\centerline{\hspace*{-1mm}\psfig{file=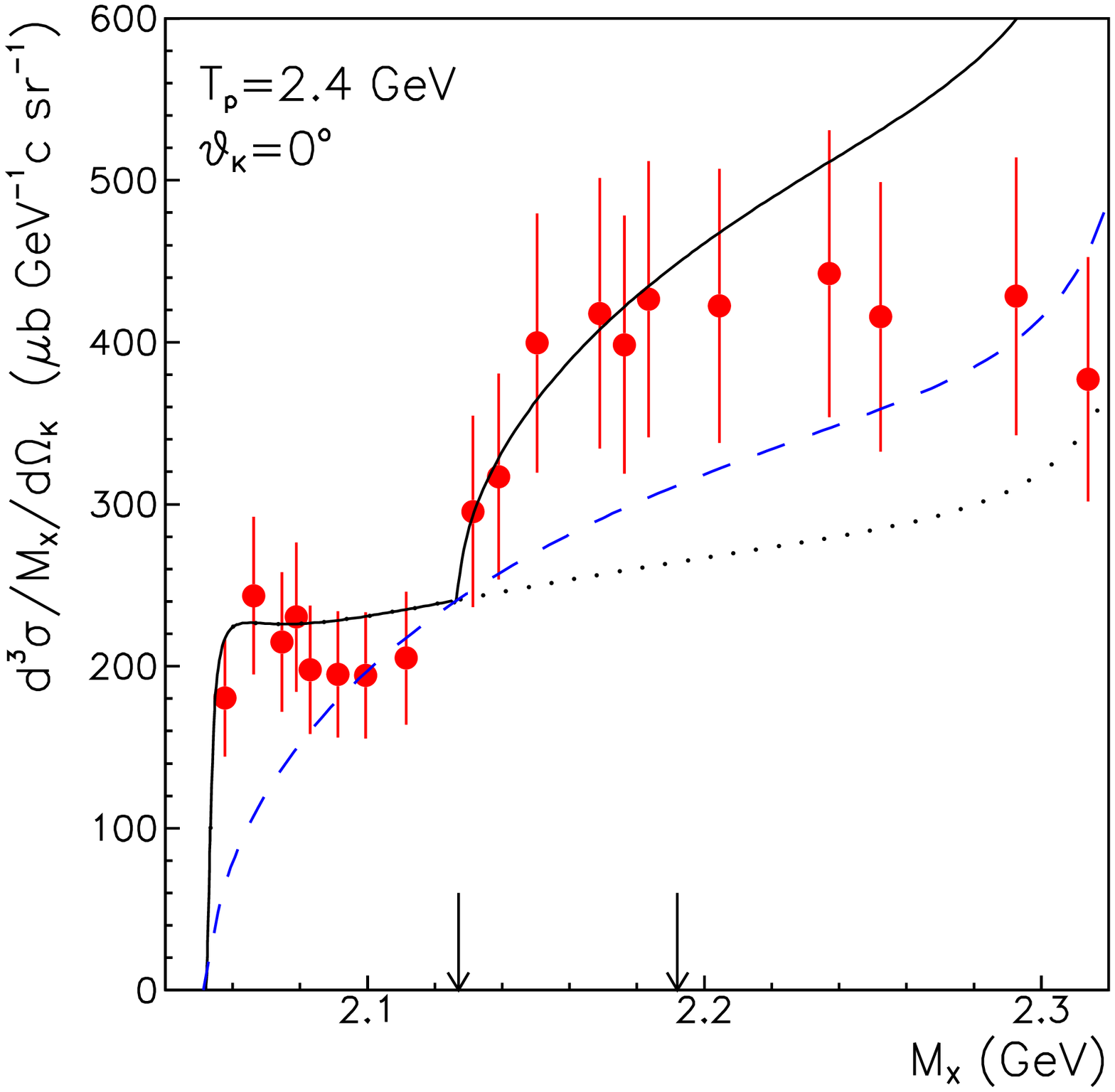,width=7.8cm,height=7cm}
\hspace*{-6mm}\psfig{file=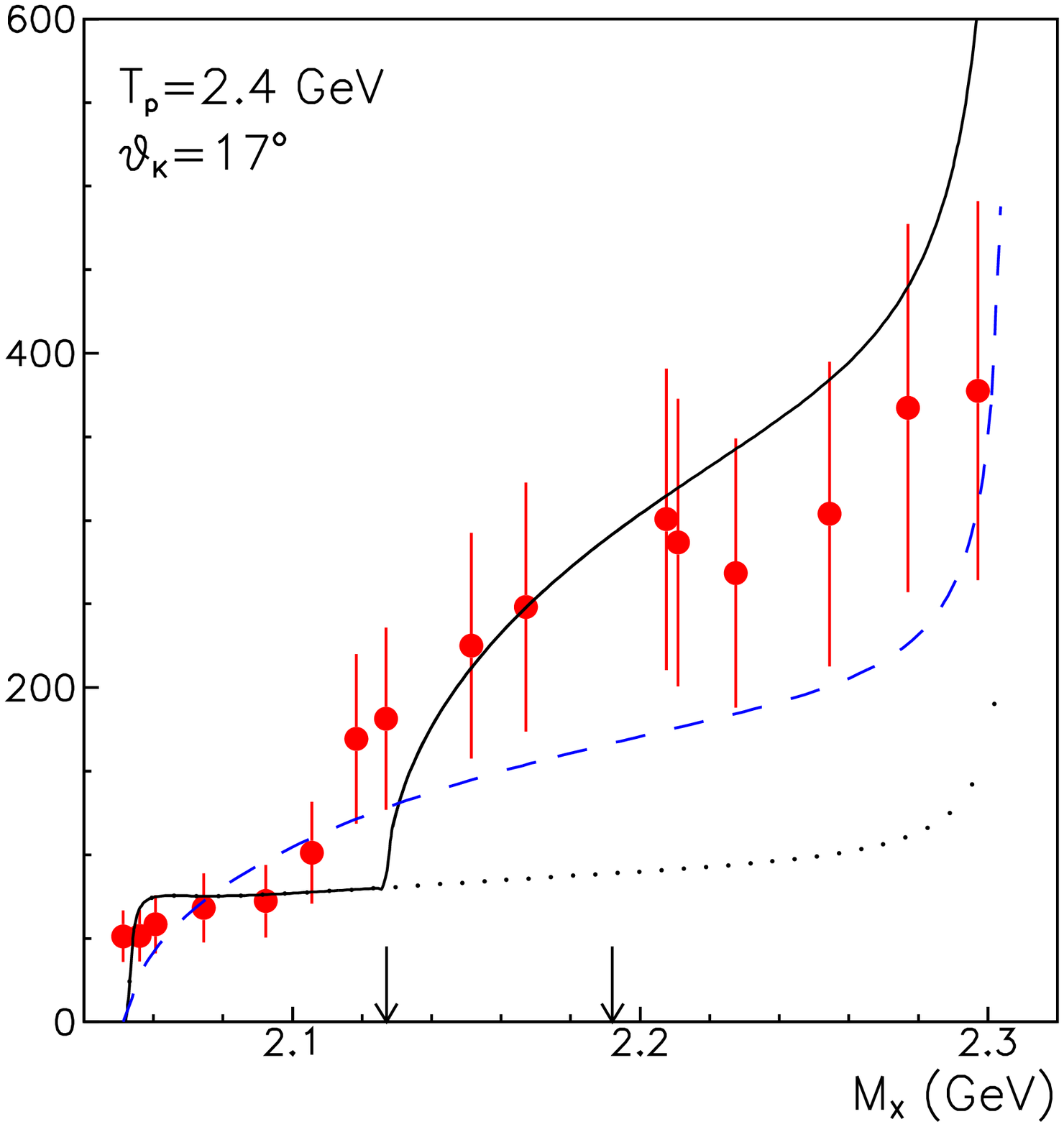,width=7.8cm,height=7cm}}
\vspace*{-3mm}
\caption{
Experimental missing mass spectra from the
$pp{\to}K^+X$ reaction~\cite{Reed} at the proton beam energy of
$T_p{=}$2.4 GeV and at different kaon production angles $\theta_K$.
Same description of curves as in Fig.~\ref{kapro6e}.} 
\label{kapro6b}
\end{figure*}

Fig.~\ref{kapro6b} shows the $M_X$-spectra 
from Ref.~\cite{Reed} at the proton beam energy of
$T_p{=}$2.4 GeV and kaon production angles of $\theta_K{=}0^o$
and 17$^o$. The notation for the lines are the same as in
Fig.~\ref{kapro6e}. Here the $K^+\Sigma{N}$ threshold is
hardly visible in the data, especially at the larger angle, 
and accordingly there are huge differences in the extracted
cross sections between the scenarios with and without $\Lambda p$
FSI, cf. Tables~\ref{TAB1a} and \ref{TAB1}. Ultimately, this is
also reflected in the large error bars for the extracted value
of $\sigma_\Sigma^\theta$. Please recall that the $\Sigma$
contributions are fitted to the invariant mass spectrum in 
the range $m_\Sigma{+}m_N \le{M_X}{\le}m_\Lambda{+}m_N{+}m_\pi$,
i.e. between the arrows shown in the figures, which explains
why the corresponding curves are in line with the data in that
region but deviate from those at higher $M_X$ values. 

\begin{figure*}[t]
\vspace*{1mm}
\centerline{\hspace*{-1mm}\psfig{file=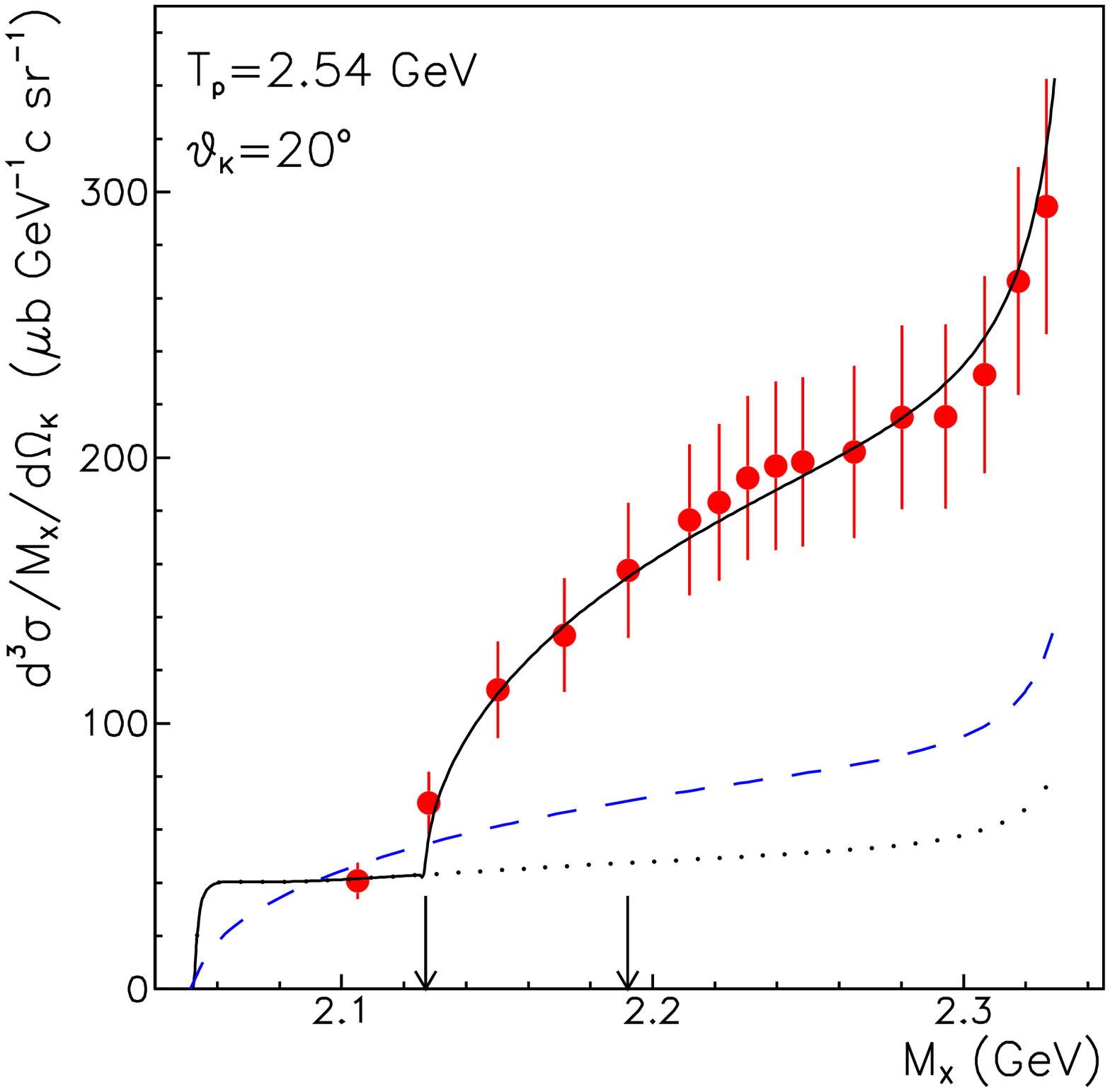,width=5.5cm,height=7.cm}
\hspace*{-5mm}\psfig{file=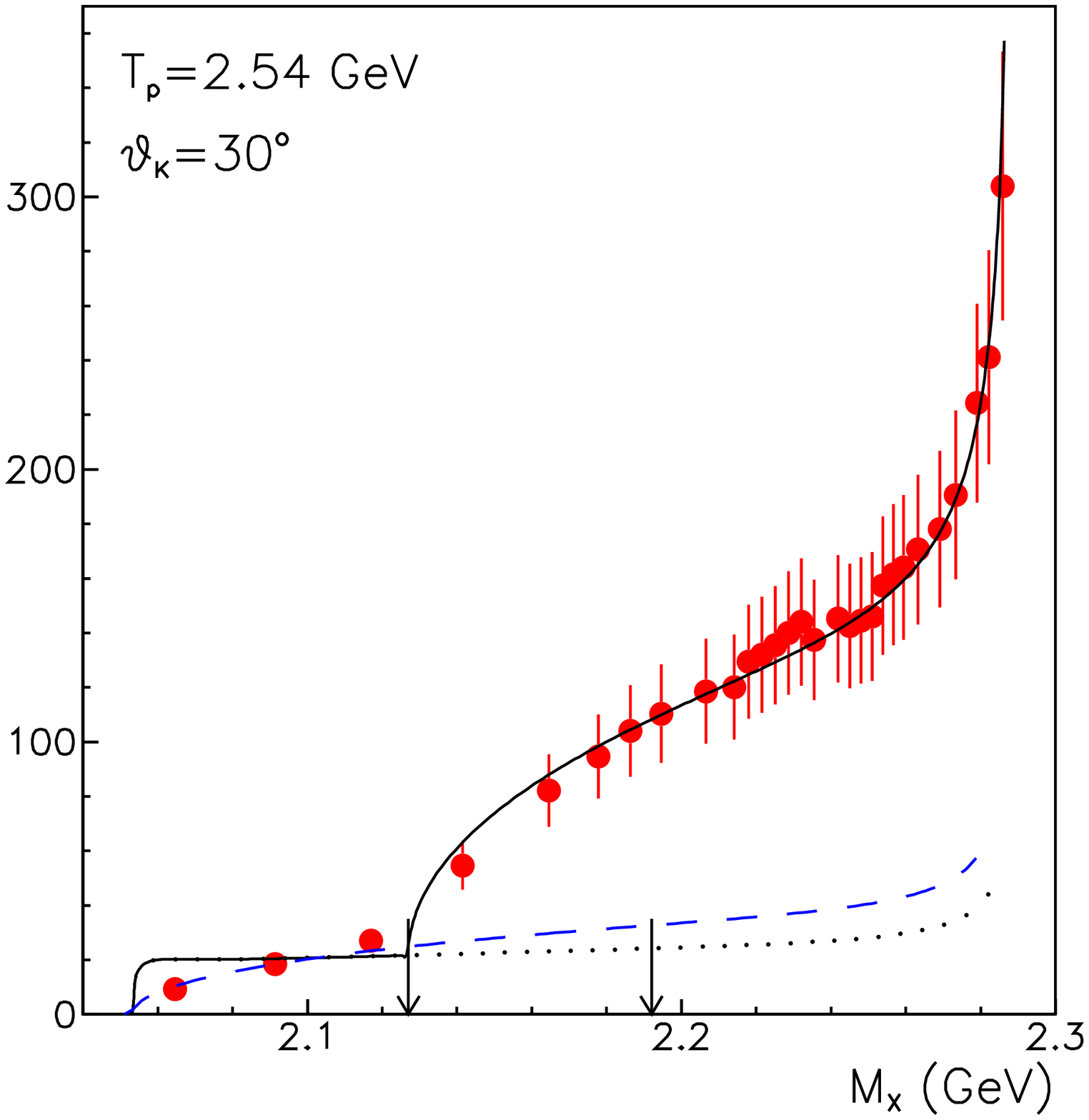,width=5.5cm,height=7.cm}\hspace*{-5mm}
\psfig{file=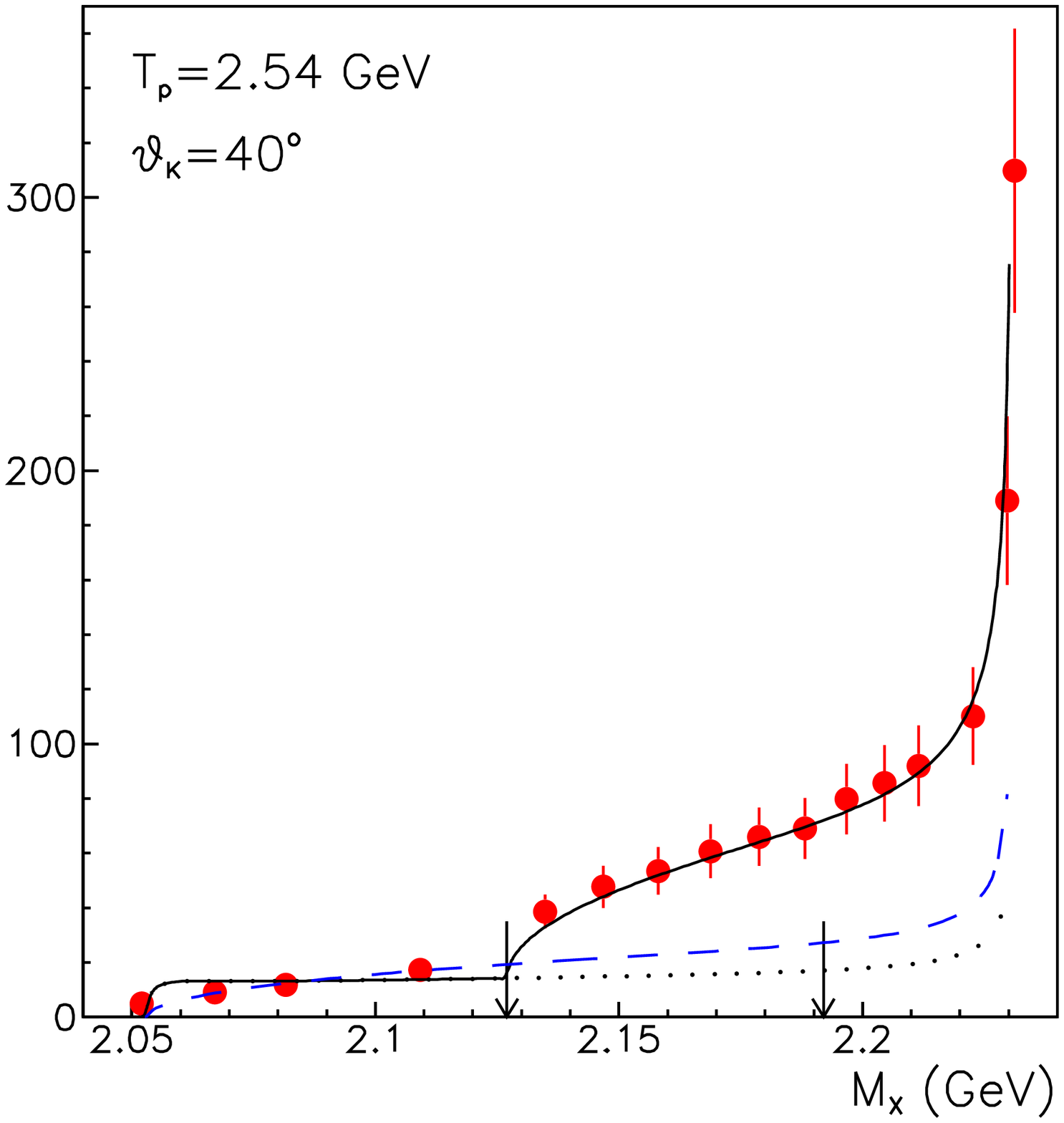,width=5.5cm,height=7.cm}}
\vspace*{-3mm}
\caption{Experimental missing mass spectra from the
$pp{\to}K^+X$ reaction~\cite{Hogan} at the proton beam energy of
$T_p{=}$2.54 GeV and at different kaon production angles $\theta_K$.
Same description of curves as in Fig.~\ref{kapro6e}.} 
\label{kapro6a}
\end{figure*}

Fig.~\ref{kapro6a} shows the missing mass spectra 
from Ref.~\cite{Hogan} at the proton beam energy of
$T_p{=}$2.54 GeV and kaon production angles of $\theta_K{=}20^o$, 30$^o$ 
and 40$^o$. There are only few points below the $K^+\Sigma{N}$
threshold and from those it is hard to see whether there is actually an 
enhancement due to the $\Lambda{p}$ FSI. 
Note that at this specific energy the description of the missing mass spectra 
for large $K^+$-meson production angles is very good, in particular, also 
for the data points above the $K^+\Lambda{N}\pi$ threshold. Thus, there
seems to be not much room for contributions from the reaction channel with 
an additional pion.

\begin{figure*}[t]
\vspace*{0mm}
\centerline{\hspace*{-1mm}\psfig{file=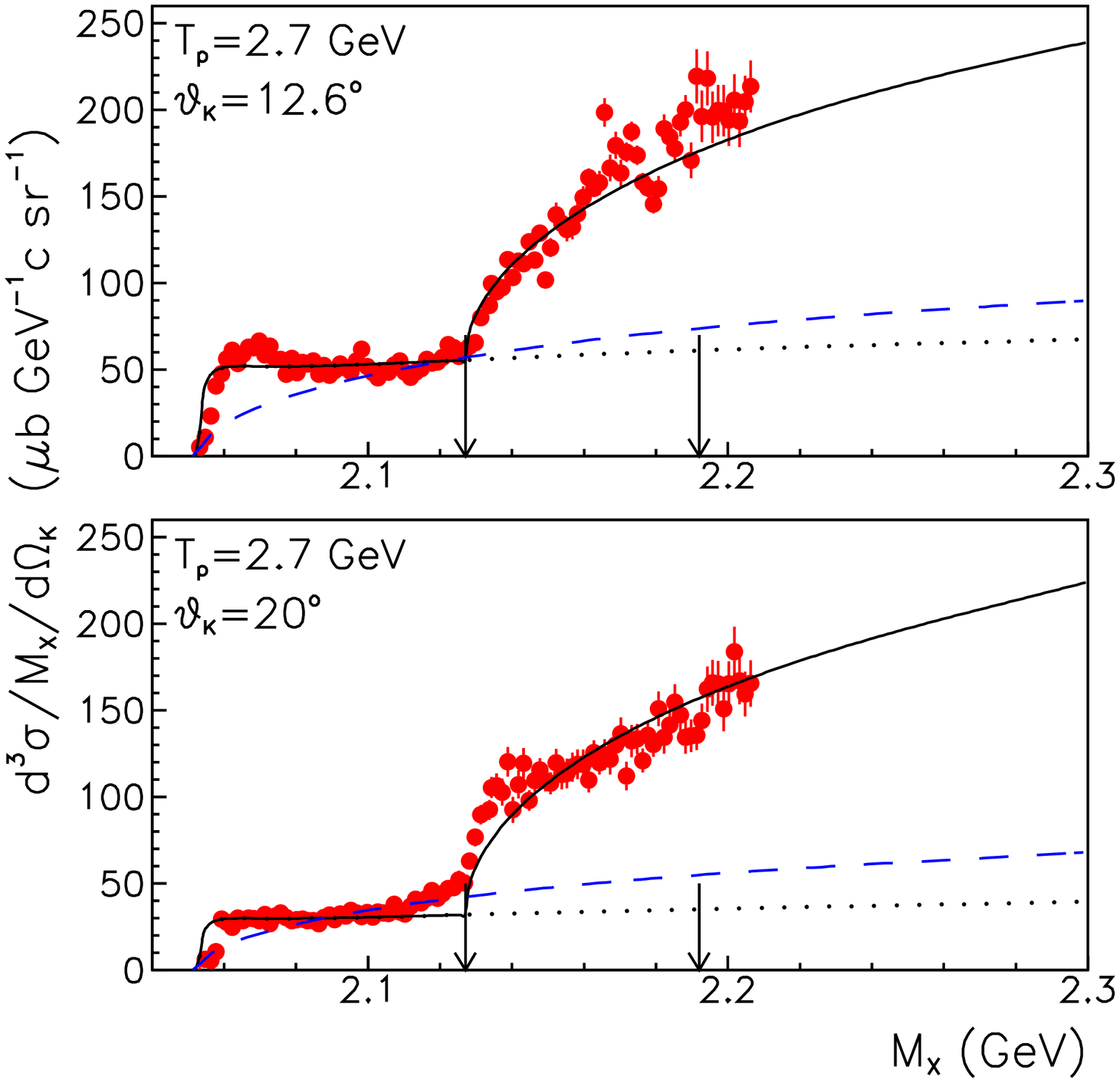,width=7.cm,height=10.1cm}
\hspace*{-6mm}\psfig{file=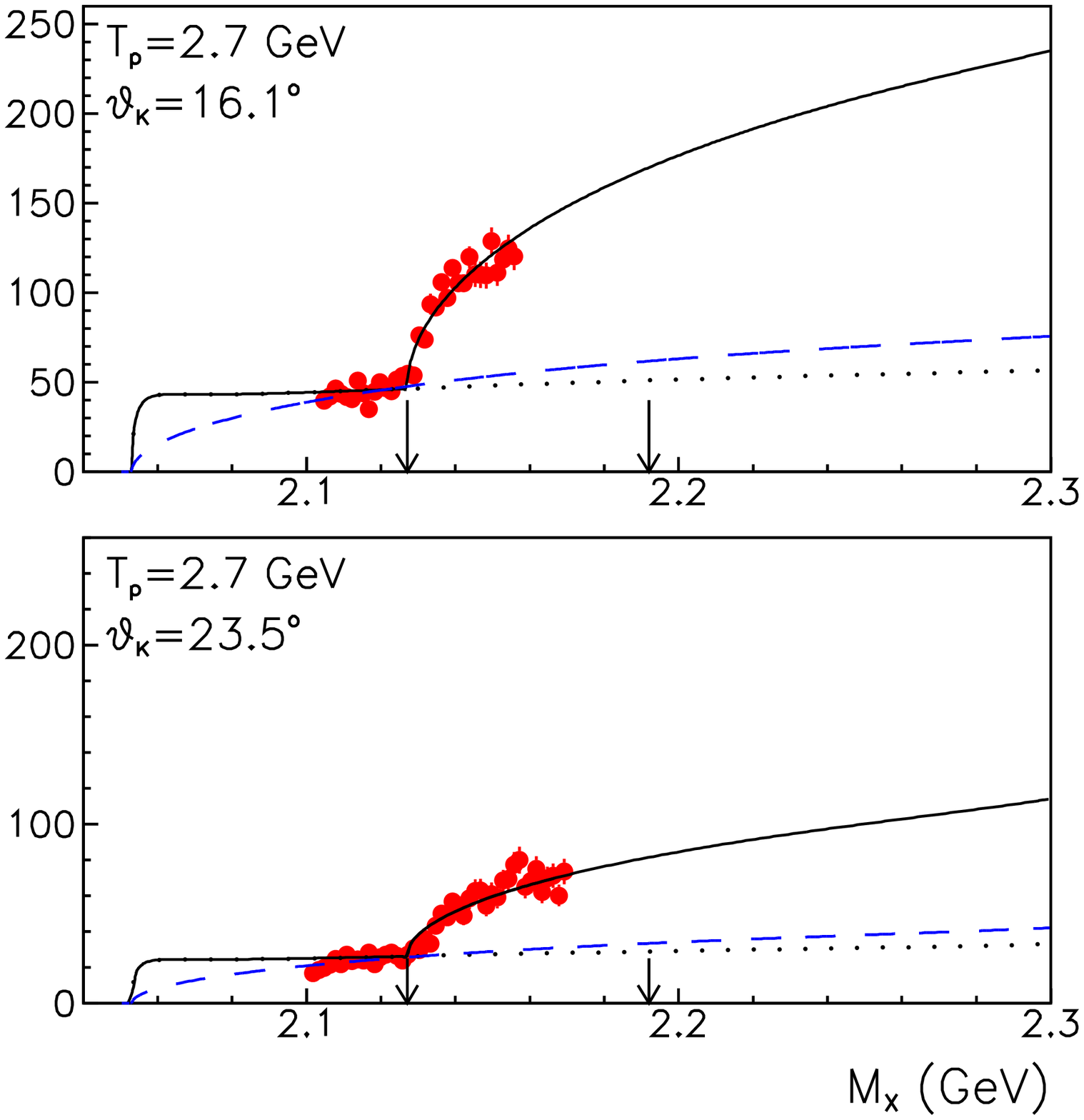,width=7.cm,height=10.1cm}}
\vspace*{-3mm}
\caption{Experimental missing mass spectra from the
$pp{\to}K^+X$ reaction~\cite{Siebert} at the proton beam energy of
$T_p{=}$2.7 GeV and different kaon production angles $\theta_K$.
Same description of curves as in Fig.~\ref{kapro6e}.} 
\label{kapro6d}
\end{figure*}

The missing mass spectra from 
Ref.~\cite{Siebert} at the proton beam energy of
$T_p{=}$2.7 GeV and kaon production angles of $\theta_K = 12.6^o$, 16.1$^o$, 
20$^o$ and 23.5$^o$ are shown in Fig.~\ref{kapro6d}. The data at
$\theta_K{=}12.6^o$ and $\theta_K{=}20^o$ indicate an enhancement due
to the $\Lambda{p}$ FSI and are well reproduced.

\begin{figure*}[t]
\vspace*{2mm}
\centerline{\hspace*{-1mm}\psfig{file=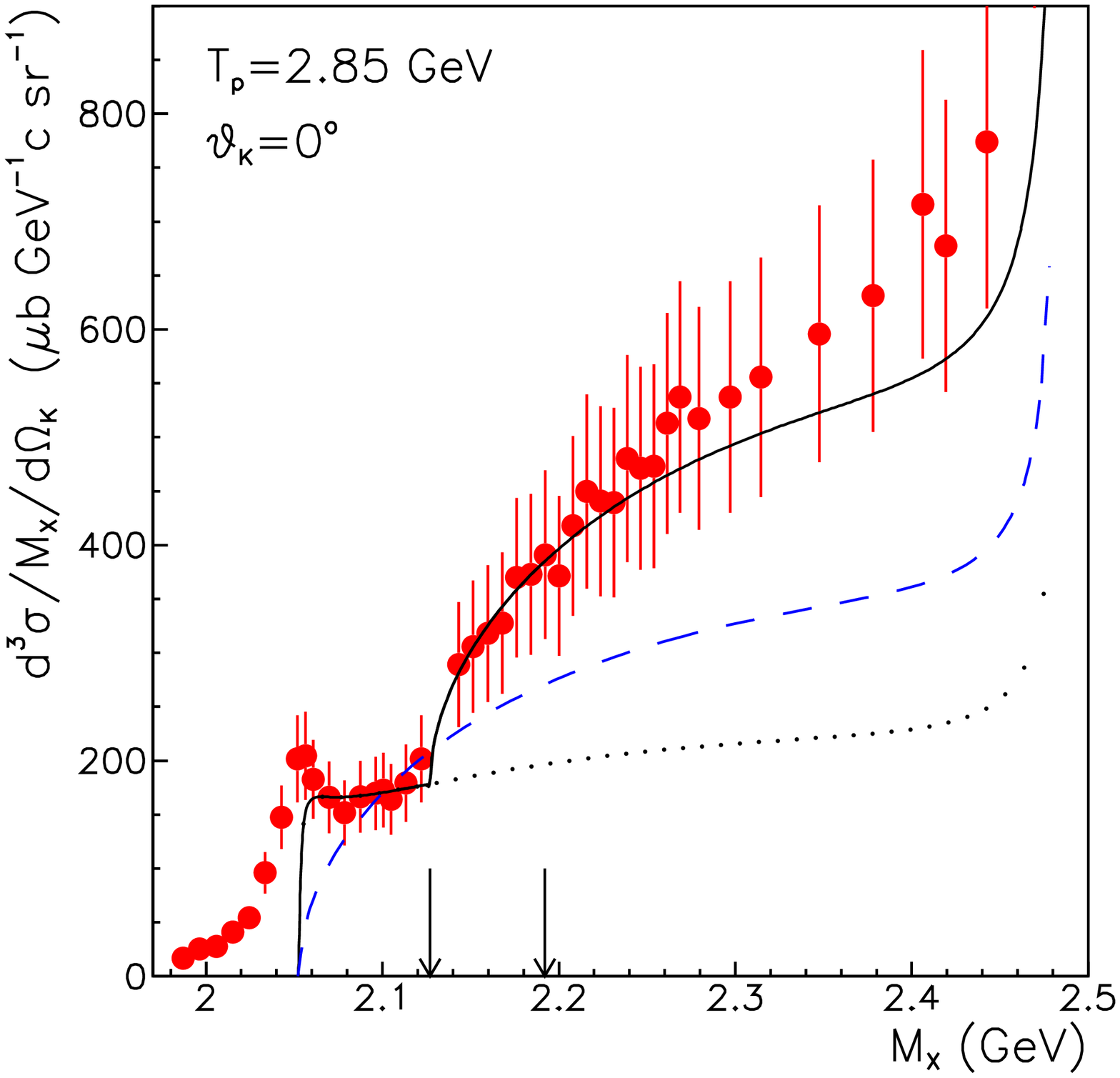,width=5.5cm,height=7.cm}
\hspace*{-5mm}\psfig{file=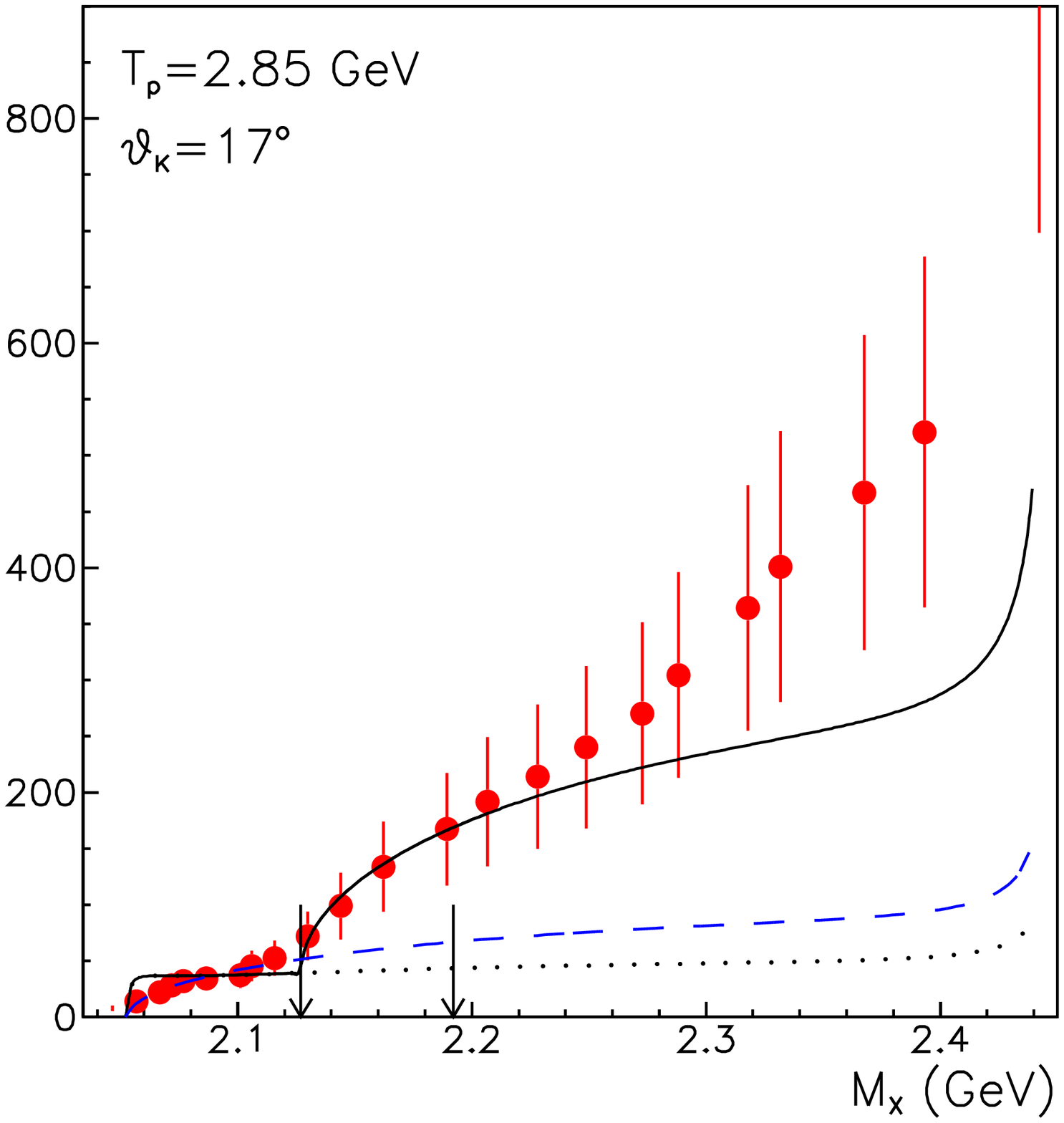,width=5.5cm,height=7.cm}\hspace*{-5mm}
\psfig{file=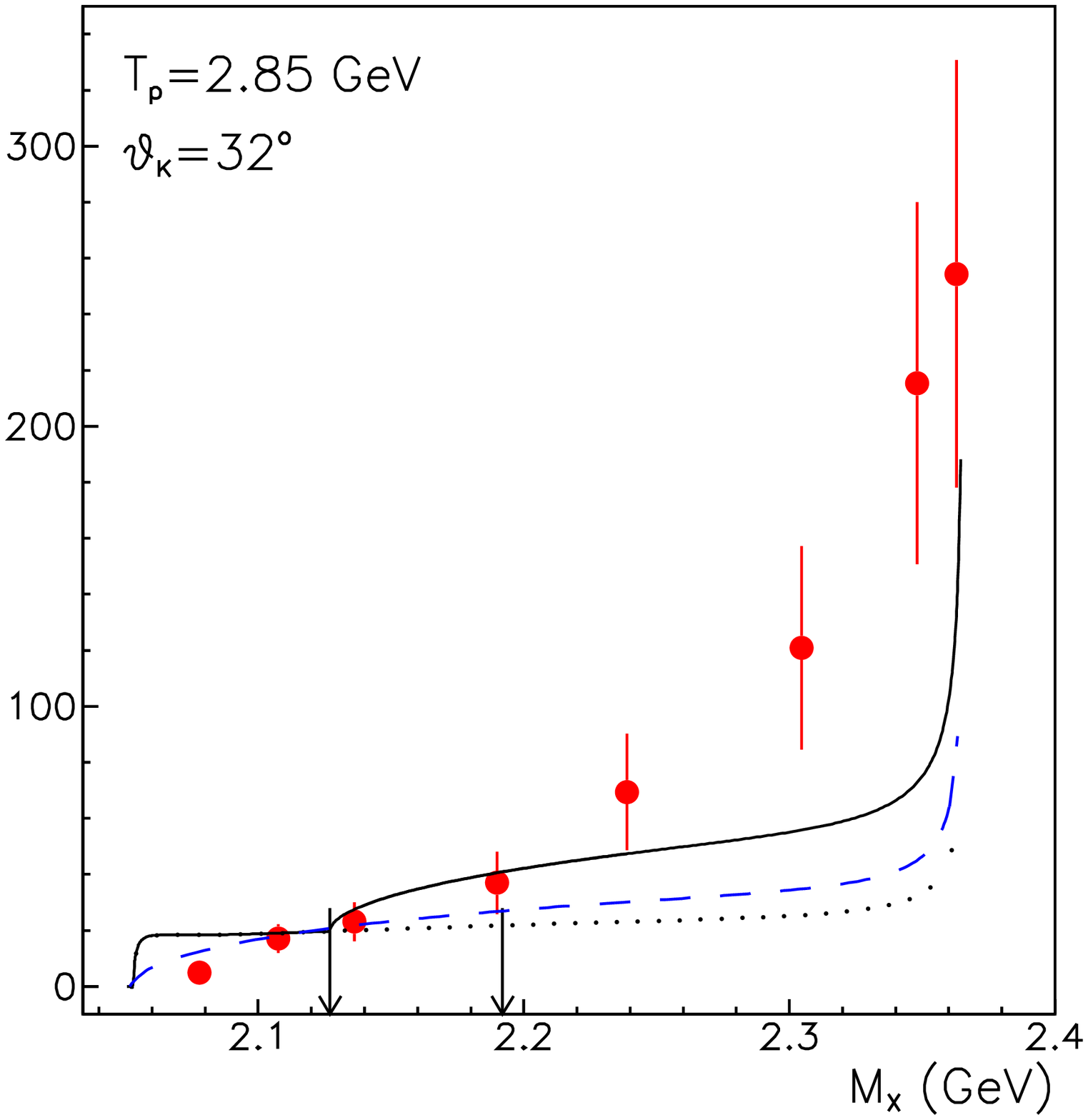,width=5.5cm,height=7.cm}}
\vspace*{-3mm}
\caption{The same as in Fig.~\ref{kapro6e} for the proton beam energy
of $T_p$=2.85 GeV. The data are from Ref.~\cite{Reed}.} 
\label{kapro6f}
\end{figure*}

Fig.~\ref{kapro6f} shows the missing mass spectra 
from Ref.~\cite{Reed} at the proton beam energy of
$T_p{=}$2.85 GeV and kaon production angles of $\theta_K{=}0^o$, 17$^o$ 
and 32$^o$. The $M_X$-spectrum at $\theta_K$=0$^o$ indicates a very
strong enhancement due to the $\Lambda{p}$ FSI. On the other hand, 
it is somewhat disturbing that there are also data points below 
the $K^+\Lambda{p}$ threshold, 
i.e. outside of the kinematically allowed region. As stated in Ref.~\cite{Reed},
this could be due to the momentum resolution of the measurement. 
In any case, those questionable events do not influence the extraction of
the $\Sigma$ production cross section from the data.

%%%%%%%%%%%%%%%%%%%%%%%%%%%%%%%%%%%%%%%%%%%%%%%%%%%%%%%%%%%%%%%%%%%%%%%%%%%%%%%%%%%%
\section{Results and discussion}

From the results collected in Tables~\ref{TAB1a} and \ref{TAB1} 
one can see that the fit to some of the data yielded
a rather small $\chi^2$, reflecting the large statistical and
systematical uncertainties of the experiments. The fit to the data at 
$T_p{=}2.7$~GeV, on the other hand, leads to a rather large $\chi^2$ because
we assume that the missing mass spectra are smooth and, therefore, we 
cannot describe the large fluctuation of the data for which very small 
statistical errors are given, as can be seen in Fig.~\ref{kapro6d}.

\begin{figure*}[t]
\vspace*{4mm}
\centerline{\hspace*{-1mm}\psfig{file=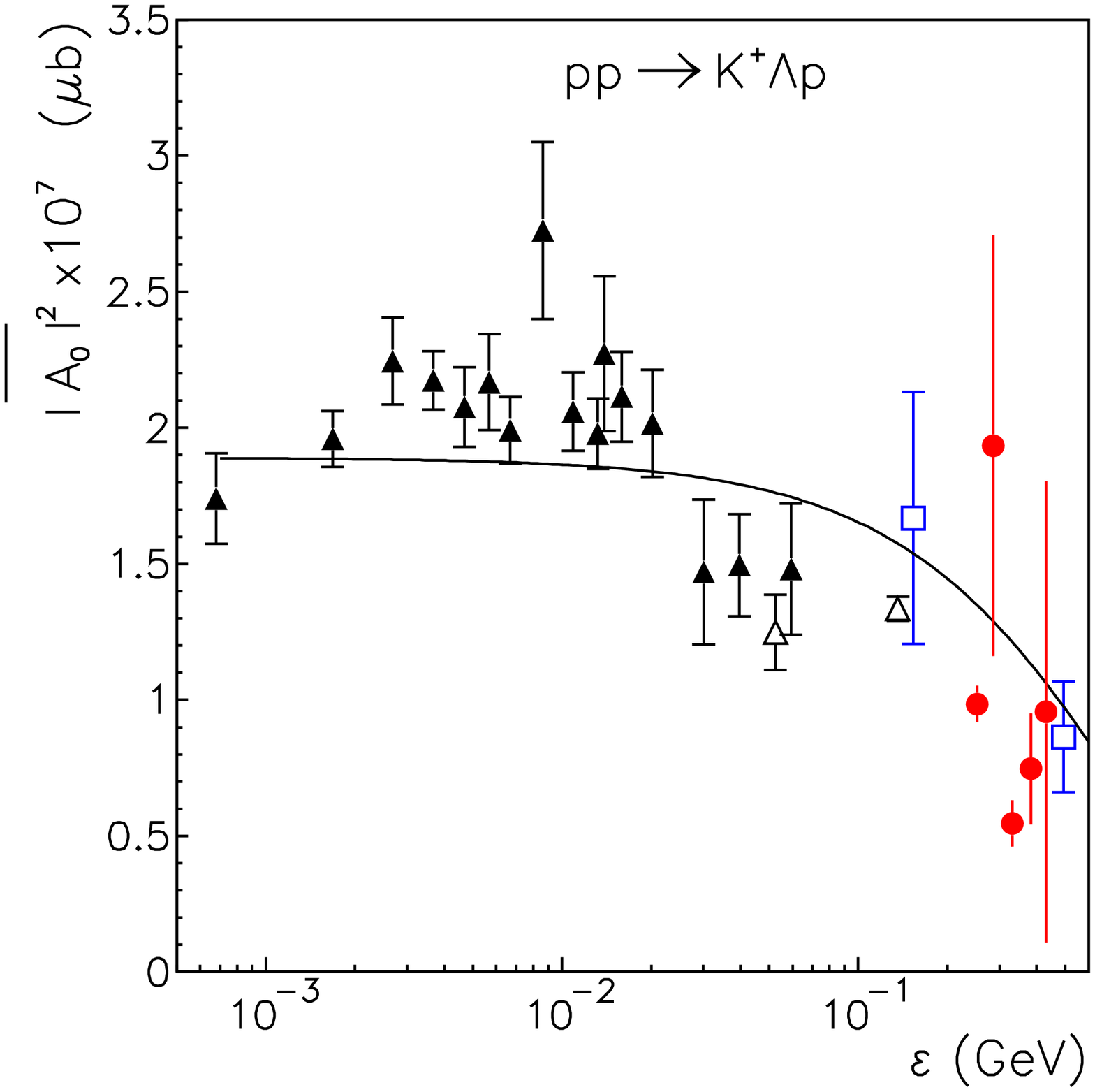,width=7.5cm,height=7.cm}\hspace*
{ -2mm}\psfig{file=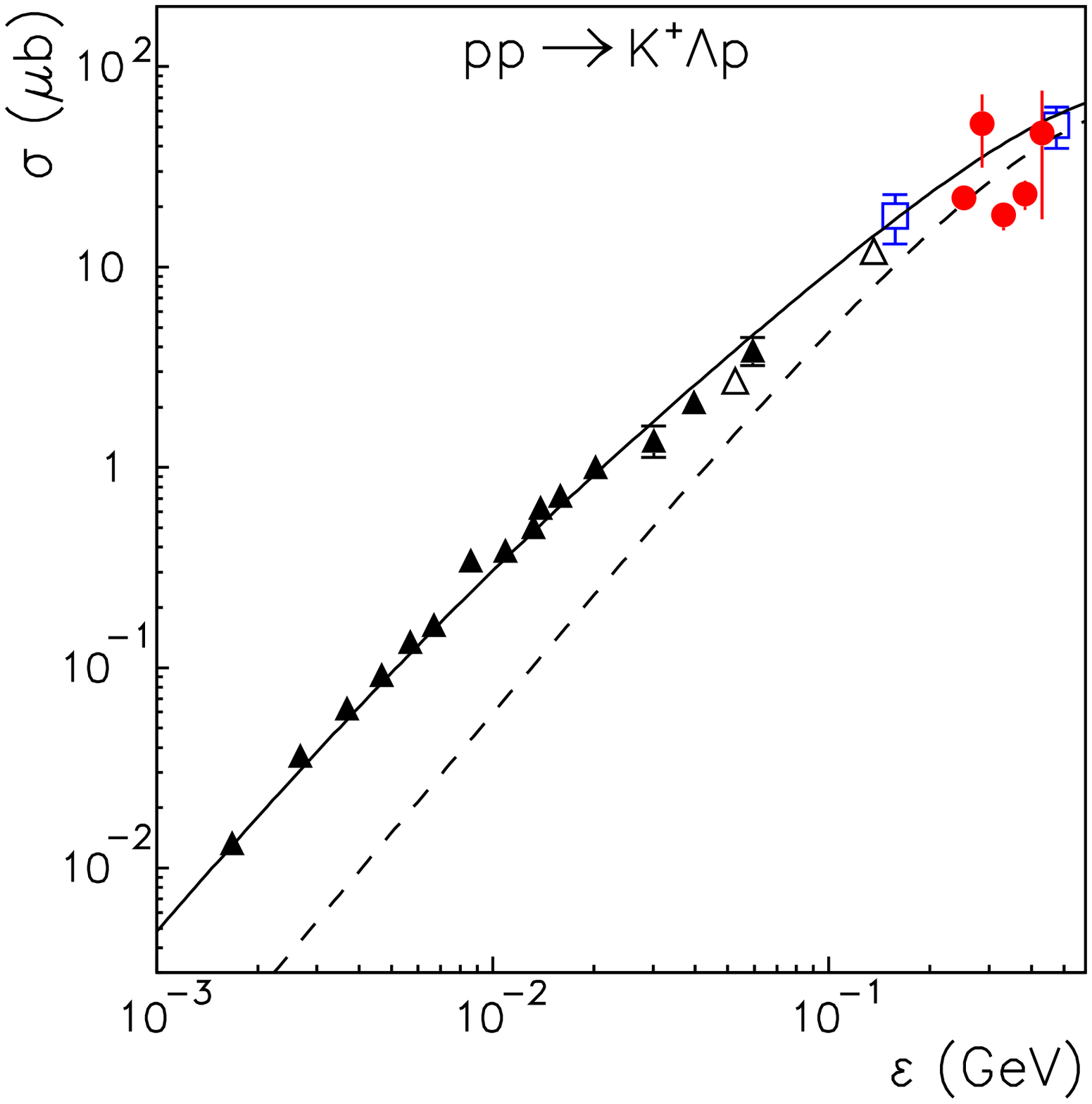,width=7.5cm,height=7.cm}}
\vspace*{-3mm}
\caption{The $pp{\to}K^+\Lambda{p}$ total reaction cross section (right) and 
corresponding squared reaction amplitude $\overline{|{\cal A}_0|^2}$ (left),
extracted via Eq.~(\ref{eval1}), as a function of the excess energy. 
The solid triangles are data from
COSY-11~\cite{Balewski1,Sewerin1,Kowina1}, the open triangles are from
COSY-TOF~\cite{Bilger1} while the open squares show data taken from
Ref.~\cite{Baldini}. The dashed and solid lines at the right side
show the result of Eq.~(\ref{eval1}) without (i.e. $\kappa$=1) and with
$\Lambda{p}$ FSI, respectively, with $\overline{|{\cal A}_0|^2}$ given 
by Eq.~(\ref{par2}). The solid line at the left side shows the 
parameterization Eq.~(\ref{par2}). The circles are results obtained 
from the missing-mass spectra analysis of the present paper.
} 
\label{kapro4}
\end{figure*}

Let us first comment on the $\xi$ factor, the indicator for the 
$\theta_K$-angular dependence of the $pp{\to}K^+\Lambda{p}$
reaction amplitude. The spectra available at $\theta_K{=}0^o$
indicate a large $\xi$ and, therefore, a strong forward peaking of 
$|{\cal A}_0|$. All other data show a smooth $\theta_K$-dependence
within the range $8.3^o{\le}\theta_K{\le}32^o$. In general, for
$\theta_K{\neq}0^o$ the factor $\xi$ is about $0.4{-}0.8$ so that the 
corresponding angle-averaged amplitude is smaller than the reference 
amplitude for the $pp{\to}K^+\Lambda{p}$ reaction computed from 
Eq.~(\ref{par2}), which represents a fit to the measured total reaction 
cross section. Corresponding results are displayed in the left panel
of Fig.~\ref{kapro4}. We want
to point out, however, that for excess energies $170{<}\epsilon{<}360$~MeV 
the uncertainty for $\overline{|{\cal A}_0|^2}$ as determined directly from 
available data points from bubble chamber measurements is also in the
order of a factor of $\simeq$2, cf. the squares in Fig.~\ref{kapro4}. 
Therefore, we conclude that there is a 
reasonable consistency between the squared reaction amplitudes deduced
from the missing mass and the values obtained from direct measurements. 

Based on the values for $\sigma_\Lambda^\theta$ at the same excess 
energies and for different $\theta_K$ from Table~\ref{TAB1} 
we can calculate the mean value and the standard deviation for 
the $pp{\to}K^+\Lambda{p}$ total reaction cross section, 
cf. Table \ref{TAB2a} and the right panel of Fig.~\ref{kapro4} (circles).
Although for some energies the standard deviations are very large, there 
is reasonable overall agreement between the results extracted from the 
missing mass spectra and the cross section data from direct measurements. 

\begin{table*}
\caption{\label{TAB2a}
Results for the total $pp\to K^+\Lambda{p}$ cross section ${\sigma_\Lambda}$.
The excess energy $\epsilon$ is given 
with respect to the $\Lambda$-hyperon production threshold.
Our results, extracted from the missing mass spectra,
averaged over the $\theta_K$ angles, are listed in the upper part of
the table. The error bars are computed from the corresponding standard 
deviation. 
Experimental results from direct measurements in a comparable energy
region and the corresponding references are listed below.
}
\footnotesize\rm
\begin{tabular*}{\textwidth}{@{}l*{15}{@{\extracolsep{0pt plus12pt}}c}}
\hline\noalign{\smallskip}
$\epsilon$ (MeV)  &
${\sigma_\Lambda}$ ($\mu$b)&  \\
\hline\noalign{\smallskip}
&\multicolumn{1}{c}{our evaluation} & missing mass spectrum \\
&  & from Ref. \\
\hline\noalign{\smallskip}
252 & 22.2$\pm$1.5 &  \cite{Siebert} \\
285  & 52.1$\pm$20.8 &   \cite{Reed} \\
331  & 18.2$\pm$2.9 &   \cite{Hogan} \\
383 & 23.1$\pm$3.8 &   \cite{Siebert} \\
431  & 46.6$\pm$29.3 &    \cite{Reed} \\
\noalign{\smallskip}\hline\noalign{\smallskip}
& \multicolumn{1}{c}{direct  measurements} & Ref. \\
\hline\noalign{\smallskip}
138 & 12.0$\pm$0.4 & \cite{Bilger1}\\
157 & 18$\pm$5 & \cite{Ficki}\\
431 & 51$\pm$12 & \cite{Louttit}\\
\hline\noalign{\smallskip}
\end{tabular*}
\end{table*}

\begin{table*}
\caption{\label{TAB2}
Results for the $\Sigma$ production cross sections. 
The excess energy $\epsilon$ is given 
with respect to the $\Sigma$-hyperon production threshold.
${\sigma_\Sigma}$ is the total $\Sigma$ production cross 
section extracted from the missing mass spectra,
averaged over the $\theta_K$ angles. The error bars
are computed from the corresponding standard deviation. 
The cross section for $\sigma(K^+\Sigma^0p)$ was obtained via 
Eq.~(\ref{eval1}), utilizing the parametrization of the cross data 
for that channel from direct measurements as given in Eq.~(\ref{par1}). 
The cross section for $\sigma(K^+\Sigma^+n)$ is identified with the 
difference between ${\sigma_\Sigma}$ and $\sigma(K^+\Sigma^0p)$.
Listed are also experimental results from direct measurements of the 
$pp{\to}K^+\Sigma^0p$ and $pp{\to}K^+\Sigma^+n$ channels and the 
corresponding references. Note that $\sigma(K^+\Sigma^0p)$ at
$\epsilon=727.6$ MeV is taken from Ref.~\cite{Bierman} and those
at 13 and 60 MeV are taken from Refs.~\cite{Sewerin1} (preprint)
and \cite{Kowina1}, respectively. 
}
\footnotesize\rm
\begin{tabular*}{\textwidth}{@{}l*{15}{@{\extracolsep{0pt plus12pt}}c}}
\hline\noalign{\smallskip}
$\epsilon$ (MeV) &  ${\sigma_\Sigma}$ ($\mu$b) &
$\sigma(K^+\Sigma^0p)$ ($\mu$b) & $\sigma(K^+\Sigma^+n)$ ($\mu$b) & \\
\hline\noalign{\smallskip}
&\multicolumn{3}{c}{our evaluation} & missing mass spectrum \\
\noalign{\smallskip}\hline\noalign{\smallskip}
& & & & from Ref. \\
178 & 17.7$\pm$2.3 & 4.0$\pm$0.3 & 13.7$\pm$2.3& \cite{Siebert} \\
212 & 43.4$\pm$7.3 & 5.2$\pm$0.5 & 38.2$\pm$7.3& \cite{Reed} \\
258 & 43.2$\pm$1.8 & 6.9$\pm$0.7 & 36.3$\pm$1.9& \cite{Hogan} \\
309 & 48.8$\pm$ 11.8 & 8.9$\pm$1.0 & 39.9$\pm$11.8& \cite{Siebert} \\
357 & 58.6$\pm$27.6 & 10.6$\pm$1.3 & 48.6$\pm$27.6& \cite{Reed} \\
\hline\noalign{\smallskip}
& \multicolumn{3}{c}{direct  measurements} & Ref. \\
\hline\noalign{\smallskip}
13 & & 0.020$\pm$0.003 & 4.56$\pm$0.94$\pm$2.7 & \cite{Rozek} \\
60 & & 0.482$\pm$0.144 & 44.8$\pm$10.7$\pm$15.2 & \cite{Rozek} \\
357 & & 13$\pm$7 & 47$\pm$13 & \cite{Louttit} \\
728 & & 25$\pm$3 & 48.1$\pm$3.5 & \cite{Sondhi} \\
849 & & 27$\pm$4 & 85$\pm$12& \cite{Dunwoodie}\\
1006 & & 17$^{+4}_{-2}$& 57$\pm$7 & \cite{Chinowsky}\\
1156 & & 25$\pm$3 & 85$\pm$11 & \cite{Dunwoodie}\\
\hline\noalign{\smallskip}
\end{tabular*}
\end{table*}

The sum of the $pp{\to}K^+\Sigma^0{p}$ and $pp{\to}K^+\Sigma^+{n}$
cross sections ($\sigma_\Sigma^\theta$), extracted from the missing
spectra, shows a more significant angular dependence only at the
highest excess energy of $\epsilon$=356.8~MeV, cf. Table~\ref{TAB1}. 
For the other energies
the dependence of $\sigma_\Sigma^\theta$ on the $K^+$-meson production 
angle is relatively smooth, at least within the uncertainties of the 
extracted values. 
From Table~\ref{TAB1} one can also see that neglecting the $\Lambda{p}$
FSI in the extraction procedure yields results for $\sigma_\Sigma^\theta$ 
which are, in general, somewhat smaller. But it is reassuring to see that
in most of the cases the error bars obtained for the $\sigma_\Sigma^\theta$'s
with and without inclusion of the $\Lambda{p}$ FSI overlap, indicating that
the results are not too sensitive to the specific subtraction prescription.
There are only a few cases where there are indeed dramatic differences between
the values for the two considered options. 
As already said, we consider the extrapolation based on the fit that includes 
the $\Lambda{p}$ as much more reliable and, therefore, we consider the $\Sigma$ 
cross sections deduced from that fit as our definitive results. 
Averaging over the $\sigma_\Sigma^\theta$ values extracted at different $K^+$-meson 
angles we can now evaluate the $\Sigma$ production cross section,
${\sigma_\Sigma}$, and determine the standard deviation for the
results obtained from the fit to the $M_X$-spectra. The corresponding values,
for the case where the $\Lambda{p}$ FSI was taken into account, are listed 
in Table~\ref{TAB2}. 

\begin{figure*}[t]
\vspace*{1mm}
\centerline{\hspace*{-1mm}\psfig{file=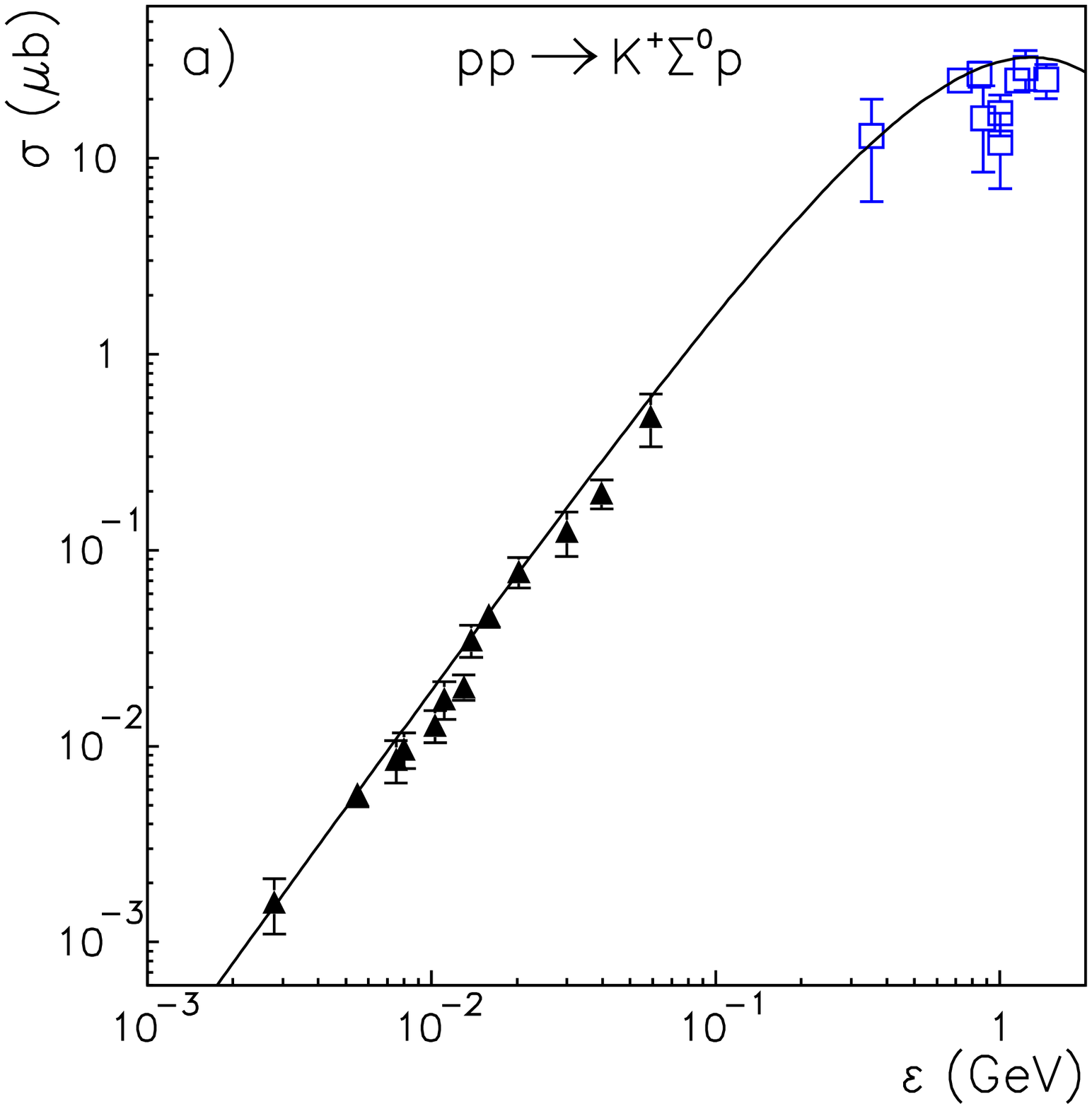,width=7.5cm,height=7.cm}
\hspace* { -2mm}\psfig{file=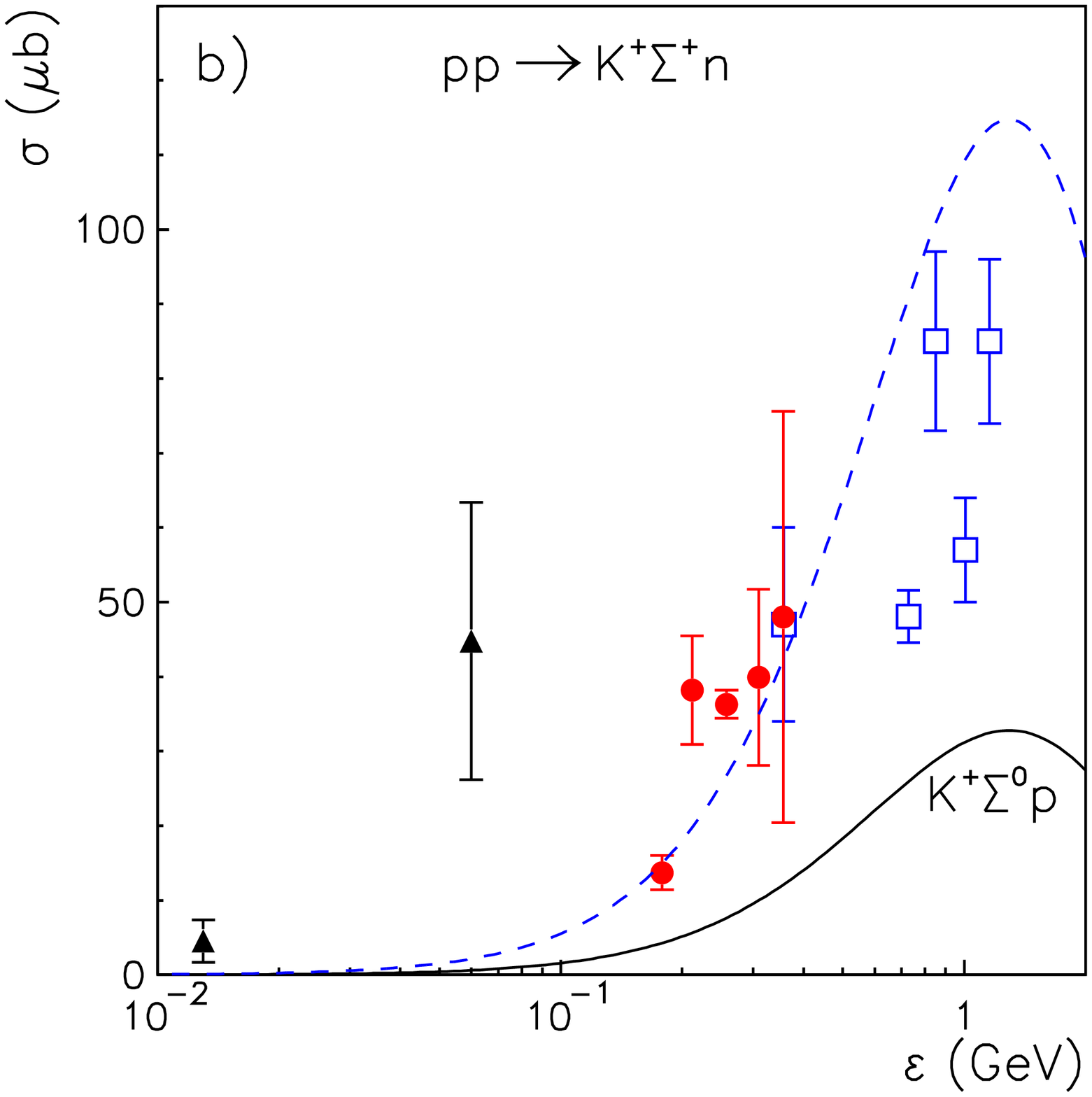,width=7.5cm,height=7.cm}}
\vspace*{-3mm}
\caption{The $pp{\to}K^+\Sigma^0{p}$ (a) and $pp{\to}K^+\Sigma^+{n}$
(b) total reaction cross sections as a function
of the excess energy. The triangles are data by the COSY-11 collaboration 
for the $K^+\Sigma^0{p}$ \cite{Sewerin1,Kowina1} and $K^+\Sigma^+{n}$  
\cite{Rozek} channels, while the squares show bubble chamber data taken from 
Refs.~\cite{Louttit,Sondhi,Dunwoodie,Chinowsky,Baldini,Bierman}. 
The circles are results for $pp{\to}K^+\Sigma^+{n}$ obtained 
from our analyis of the missing mass spectra. The solid lines in both panels 
show the result of Eq.~(\ref{eval1}) without $\Sigma^0{p}$ FSI 
(i.e. with $\kappa$=1) and with $\overline{|{\cal A}_{\Sigma^0}|^2}$ given by
Eq.~(\ref{par1}). The dashed line in (b) is the same as the 
solid line, but multiplied by a factor 3.5. } 
\label{kapro7}
\end{figure*}

In order
to deduce the $pp{\to}K^+\Sigma^+{n}$ cross section
one needs to subtract from ${\sigma_\Sigma}$ the
$pp{\to}K^+\Sigma^0{p}$ cross section. In 
Fig.~\ref{kapro7}a we show the existing data for the reaction
$pp{\to}K^+\Sigma^0{p}$ as a function of the excess
energy. The circles are from the COSY-11
Collaboration~\cite{Sewerin1,Kowina1}, while the open squares 
are bubble chamber data~\cite{Baldini}.
Presently there are no experimental results available for 
$60{<}\epsilon{<}360$~MeV, i.e. for the energy range of our
analysis. The measurements from the TOF-Collaboration that cover 
the energy range of our interest are still at the stage of being 
analysed~\cite{TOF}. Therefore, to proceed further, 
we fit the $pp{\to}K^+\Sigma^0{p}$ total reaction cross section by 
Eq.~(\ref{eval1}) with $\kappa{=}1$, i.e. by neglecting the 
$\Sigma^0p$ FSI, and with the appropriate kinematics for the
$K^+\Sigma^0{p}$ channel. The resulting squared reaction amplitude is
\begin{eqnarray}
\overline{|{\cal A}_{\Sigma^0}|^2}
=(0.61\pm 0.03)\cdot\exp[(1.34\pm 0.2)\epsilon]
\cdot 10^7  \,\,\, (\mu{\rm b}),
\label{par1}
\end{eqnarray}
with the excess energy given in GeV. Note that the 
omission of possible $\Sigma^0p$ FSI effects is in line with 
the experimental evidence for the $pp{\to}K^+\Sigma^0p$ 
channel~\cite{Sewerin1,Kowina1} -- the available data do not show 
any visible indication for such a FSI~\cite{Sibirtsev1a,Kowina1}. 
It is also in line with
the conclusions we draw from inspecting the experimental
mass spectra analysed in the present paper, which likewise
exhibit no sign for the presence of a $\Sigma N$ FSI effects. 

The parameterization (\ref{par1}) allows us to
calculate the $pp \to K^+\Sigma^0{p}$ cross section for each of the 
excess energies, where data on the $pp{\to}K^+X$ reaction exist.
The corresponding values are listed in Table~\ref{TAB2}. The last column in 
Table~\ref{TAB2} is the difference between ${\sigma_\Sigma}$ 
and the $pp{\to}K^+\Sigma^0{p}$ cross section, which we identify with 
the $pp{\to}K^+\Sigma^+{n}$ cross section. The results are also shown 
in Fig.~\ref{kapro7}b (circles). Note that a linear scale is used for 
displaying the $pp{\to}K^+\Sigma^+{n}$ cross section! 
 
First let us compare our results with the bubble chamber
data~\cite{Louttit,Sondhi,Dunwoodie,Chinowsky} 
shown by the open squares in Fig.~\ref{kapro7}b and listed also
in Table~\ref{TAB2}. The $pp{\to}K^+\Sigma^+{n}$ cross section
of 47$\pm$13 $\mu$b at $\epsilon$=357~MeV was measured 
by Louttit et al.~\cite{Louttit} and it is in agreement with our result 
at the same energy, cf. Table~\ref{TAB2} -- though, unfortunately, 
at this highest energy of our analysis there is a large uncertainty 
due to the angular dependence of the extracted cross section as can be
seen from Table~\ref{TAB1}. The $pp{\to}K^+\Sigma^+{n}$ 
data~\cite{Sondhi,Dunwoodie,Chinowsky} at higher energies indicate 
large fluctuations. We want to remark that the two
points at $\epsilon \simeq $ 0.85 and 1.1~GeV represent the 
largest reported cross sections with 85 $\mu$b (at both energies) and 
are available~\cite{Dunwoodie} only in an UCLA preprint.  
For the lowest energy where missing mass spectra are available, 
$\epsilon$=178 MeV, we deduced a cross section of $13.7\pm2.3$ $\mu b$ 
from the data. There are two so far unpublished measurements of the 
$pp{\to}K^+\Sigma^+{n}$ cross section by the TOF collaboration at 
somewhat lower energy, i.e. at beam momenta of 2.06 GeV 
($\epsilon$=98 MeV) \cite{Schoen} and 2.157 GeV 
($\epsilon$=128 MeV) \cite{Karsch}, respectively. It is worth 
mentioning that their results are roughly in line with the value we 
obtained. 
The results from the COSY-11 Collaboration at low excess energies 
have been available only recently~\cite{Rozek}. Their
$pp{\to}K^+\Sigma^+{n}$ cross section of $44.8{\pm}10.7{\pm}15.2$
at $\epsilon$=60~MeV is as large as those at high energies, 
cf. Table~\ref{TAB2} and Fig.~\ref{kapro7}. 

For illustration purposes we include the result for the
$pp \to K^+\Sigma^0{p}$ cross section also in Fig.~\ref{kapro7}b
(solid line). The dashed line shows the same cross section, but 
multiplied by a factor 3.5. 
Obviously its energy dependence is very different from 
that exhibited by the $pp{\to}K^+\Sigma^+{n}$ cross section 
if one considers {\it all} available data. However, it is 
interesting to see that the curve would be roughly in line with the 
trend of the $K^+\Sigma^+{n}$ data, including the ones obtained from 
our analysis, if one disregards the COSY-11 events
and the measurements from Refs. \cite{Sondhi,Chinowsky}. 
The data from the last two references are in clear contradiction 
to the results from Ref. \cite{Dunwoodie} anyway. On the other hand,
one has to keep in mind that the latter data were never officially 
published.

Provided that all data are indeed correct, it will be difficult to 
find plausible explanations for the drastically different behavior of
the $pp{\to}K^+\Sigma^+{n}$ cross section.
The model calculations~\cite{Laget,Sibirtsev2,Tsushima,Shyam1,Zou}
available for this reaction channel indicate that the 
energy dependence of
the cross section is similar to the one of $pp{\to}K^+\Sigma^0{p}$.
After all, the energy dependence is to a considerable part 
determined simply by phase-space factors. Evidently, the model
predictions \cite{Laget,Sibirtsev2,Tsushima,Shyam1} disagree strongly 
with the new data~\cite{Rozek} of the \hbox{COSY-11} Collaboration. 
Since those calculations 
describe the $pp{\to}K^+\Sigma^+{n}$, $pp{\to}K^+\Sigma^0{p}$ and
$pp \to K^0\Sigma^+{p}$ reactions with the same dynamical input,
additional mechanisms can be introduced only by assuming that they
contribute to the $pp{\to}K^+\Sigma^+{n}$ channel alone. 
But not even a recent study that focusses on the $pp{\to}K^+\Sigma^+{n}$ 
reaction only and invokes contributions from the $\Delta$(1620) 
resonance is able to describe the COSY-11 data satisfactorily \cite{Zou}. 
Of course, possible additional contributions could arise from the
excitation of crypto-exotic baryons, as was speculated in
Ref.~\cite{Rozek}, that then decay into the $K\Sigma$ channel. 
Such crypto-exotic baryons were discussed~\cite{Sibirtsev3} recently in
the context of the new ANKE-COSY results~\cite{Hartmann} on 
$\phi$-meson production. An indication for a possible crypto-exotic 
baryon was also reported in Ref.~\cite{Antipov}, based on an analysis 
of the $\Sigma^0K^+$ invariant mass spectrum.

Anyway, instead of embarking on further speculations we believe that 
it would be more
instructive to perform new measurements of the $pp{\to}K^+\Sigma^+{n}$ 
reaction in the near-threshold region. 
The method used in the present paper can be also applied in the analysis 
of data that can be taken at ANKE~\cite{Barsov} and HIRES~\cite{Hires}
at COSY. These experimental facilities are perfectly suited for obtaining 
$K^+$-meson spectra with high statistics and high resolution.
Such exeriments would also allow to shed light on the angular dependence 
of the reaction amplitude, which we expect to be very weak at low 
excess energies.

\section{Summary}
In the present paper we determined the sum of the
$pp{\to}K^+\Sigma^+{n}$ and $pp{\to}K^+\Sigma^0{p}$ cross sections
from inclusive $K^+$-meson momentum spectra in the 
energy range $T_p$ = 2.3 - 2.85 GeV, 
available in the literature. 
We showed that, 
after transformation of the momentum spectrum to the missing mass
($M_X$) spectrum, the contribution from the reaction channels with 
$\Lambda$ and $\Sigma$ hyperons can be \, isolated by \, inspecting \, the
data between the $K^+\Lambda{p}$, $K^+\Sigma{N}$, and $K^+\Lambda{N}\pi$
thresholds and we demonstrated that the $M_X$-spectra can be well described 
when taking into account the contributions from the $pp{\to}K^+\Lambda{p}$, 
$pp{\to}K^+\Sigma^+{n}$ and $pp{\to}K^+\Sigma^0{p}$ reactions. 
The angular dependence of the reaction amplitude was accounted for by 
fitting the $K^+$-meson spectra at different angles. Total cross 
sections were then deduced by averaging over the angles. 

As a test we first 
determined the $pp{\to}K^+\Lambda{p}$ cross sections at those 
excess energies where the invariant mass spectra are availabe. It turned 
out that the cross sections extracted by us are roughly in line with results 
from direct measurements in the same energy region. 

Utilizing available information on the $pp{\to}K^+\Sigma^0{p}$ 
cross section, we then deduced total cross sections for the 
$pp \to K^+\Sigma^+{n}$ channel. The obtained results were discussed
and compared with existing data from direct measurements. 
At the specific
energy $T_p$ = 2.85 GeV there is also a data point from a bubble 
chamber measurement~\cite{Louttit} and it was reassuring to see that 
our result is compatible with that experiment. 
The cross section obtained for $T_p$ = 2.3 GeV is with $13.7\pm2.3$ $\mu b$ 
considerably smaller than the value found in a recent experiment by
the COSY-11 Collaboration at a somewhat lower beam energy
Thus, our new cross section values, together with the already available data, 
indicate that the energy dependence of the cross section for the reaction 
$pp{\to}K^+\Sigma^+{n}$ could differ drastically from that of the 
$pp{\to}K^+\Sigma^0{p}$ channel. This would be certainly 
rather surprising. Apparently, further experiments are necessary to 
confirm this unusual behaviour. If such experiments indeed corroborate 
the present findings then it is likely that peculiar and potentially
exotic mechanisms play a role in the reaction $pp{\to}K^+\Sigma^+{n}$. 

\subsection*{Acknowledgments}
This work was partially supported by Deutsche
Forschungsgemeinschaft through funds provided to the SFB/TR 16
``Subnuclear Structure of Matter''. This research is part of the \, EU
Integrated \, Infrastructure \, Initiative Hadron Physics Project under
contract number RII3-CT-2004-506078. A.S. acknowledges support by the
COSY FFE grant No. 41760632 (COSY-085) and the JLab grant SURA-06-C0452.

\appendix 
%%%%%%%%%%%%%%% APPENDIX %%%%%%%%%%%%%%%%%%%%%%%%%%%%%%%%
\section{Treatment of the final-state interaction}
\label{App} 
In the present work we take into account effects of the final-state 
interaction in the $\Lambda p$ channel. Following standard arguments
\cite{Watson,Migdal} we 
assume that the reaction amplitude ${\cal A}$ can be factorized into 
a practically momentum and energy independent elementary production 
amplitude ${\cal A}_0$ and an FSI factor:
\begin{eqnarray}
{\cal A} \approx {\cal A}_0 \times {\cal A}_{\Lambda p} \ .
\label{WM}
\end{eqnarray}
The FSI effects are then taken into acount within the Jost function 
approach
\begin{eqnarray}
|{\cal A}_{\Lambda p}|^2 \approx \frac{q^2+\alpha^2}{q^2+\beta^2} \ ,
\label{jost}
\end{eqnarray}
where the momentum $q$ is given by
\begin{eqnarray}
q=\frac{\lambda^{1/2}(s_Q,m_\Lambda^2,m_p^2)}{2\sqrt{s_Q}} \ ,
\end{eqnarray}
and the parameters $\alpha$ and $\beta$ were taken as 
\begin{eqnarray}
\alpha=-72.3 \,\, \rm{MeV} \,,\,\, \,\,\, \beta = 212.7 \,\, \rm{MeV}.
\label{paral}
\end{eqnarray}
These parameters are related to the scattering length $a$ and effective range 
$r$ of the $\Lambda{p}$ interaction \cite{Sibirtsev1}. To be specific, they
correspond to the values $a$ = -1.8 fm and $r$ = 2.8 fm. The parameters of the 
$\Lambda{p}$ FSI that we use here were obtained in Refs.~\cite{Sibirtsev1,Sibirtsev1a}
from a global phenomenological analysis of all available data on the 
reaction $pp{\to}K^+\Lambda{p}$. But, one should keep in mind that the 
scattering parameters are not fixed uniquely. 
Actually, we have shown in Ref.~\cite{Sibirtsev1} that a large set of
different values for the scattering length $a$ and effective range $r$ 
allows to reproduce the energy dependence of the $pp{\to}K^+\Lambda{p}$ 
cross section data. Some of these parameters coincide with results 
predicted by modern $YN$
models~\cite{Rijken,Hammer,Haidenbauer1,Henk,Rijken06}.

Based on Eqs.~(\ref{WM}-\ref{jost}) one can then write the total reaction cross 
section for the reaction $pp{\to}K^+\Lambda{p}$ in the form \cite{Sibirtsev1} 
\begin{eqnarray}
\sigma_\Lambda (\epsilon) = \frac{\Phi_3}{2^6\pi^5 
\lambda^{1/2}(s,m_p^2,m_p^2)} \, \, \,
\overline{|{\cal A}_0|^2} \,\, \kappa(\epsilon) \ .
\label{eval1}
\end{eqnarray}
Here $\overline{|{\cal A}_0|^2}$ is the (angle) averaged reaction amplitude squared
while $\kappa$ is a factor that represents the FSI effects. 
The 3-body phase space is
\begin{eqnarray}
\Phi_3 =\frac{\pi^2}{4s}\int\limits^{(\sqrt{s}-m_K)^2}_{(m_\Lambda+m_p)^2}
\!\!\lambda^{1/2}(s,s_Q,m_K^2)\,\,
\nonumber \\ \times\lambda^{1/2}(s_Q,m_\Lambda^2,m_p^2)\,
\frac{ds_Q}{s_Q} \ . 
\end{eqnarray}
In the nonrelativistic limit it reduces to 
\begin{eqnarray}
\Phi_3 \to \frac{1}{2^7\pi^2}\, \frac{\sqrt{m_K m_\Lambda m_p}}{(m_K+m_\Lambda+m_p)^{3/2}}
\,\, \epsilon^2 \ .
\end{eqnarray}
The non-relativistic form is a good approximation 
for the $pp \to K^+\Lambda{p}$ reaction up to excess energies  
of $\epsilon{\simeq}$1~GeV and we use it in the present investigation. 
Also, in this case the factor $\kappa$
can be computed analytically for the Jost function approach (\ref{jost}),
and under the assumption that there is only a FSI in the $\Lambda p$ 
system~\cite{Sibirtsev1}. It amounts to 
\begin{eqnarray}
\kappa(\epsilon) =
1+\frac{4\beta^2-4\alpha^2}{(-\alpha+\sqrt{\alpha^2+2\mu\epsilon})^2} \ ,
\label{factor}
\end{eqnarray}
where $\mu$ stands for the reduced mass
\begin{eqnarray}
\mu =\frac{m_\Lambda m_p}{m_\Lambda+m_p} \ , 
\end{eqnarray}
and the parameters $\alpha$ and $\beta$ for the $\Lambda{p}$ FSI given by 
Eq.~(\ref{paral}).
Note that $\kappa \equiv 1$ in case that FSI effects are neglected. 

The $pp{\to}K^+\Lambda{p}$ reaction can be directly identified 
through the detection of the final particles and, therefore, there are
already many precise cross section data available in the literature 
\cite{Balewski1,Sewerin1,Kowina1,Bilger1,Baldini}. 
These data, displayed in Fig.~\ref{kapro4} (right side), allow a 
straight-forward determination of the squared reaction amplitude 
$\overline{|{\cal A}_0|^2}$ by means of Eq.~(\ref{eval1}).
Corresponding results are shown in Fig.~\ref{kapro4} on the left hand
side. For our purposes it is also convenient to parametrize the 
experimental cross section by means of a simple function. This is 
achieved with 
\begin{eqnarray}
\overline{|{\cal A}_0|^2}=
(1.89{\pm}0.04)\cdot\exp[(1.34{\pm}0.1)\,\epsilon]
\cdot 10^7 \,\,\, (\mu{\rm b}) \ , 
\label{par2}
\end{eqnarray}
where the parameters were determined by a fit 
to the data in Fig.~\ref{kapro4} for energies $\epsilon{<}$500~MeV.
Here $\epsilon$ is the excess energy in GeV.
The corresponding curve, including also the $\Lambda{p}$ FSI via
Eq.~(\ref{factor}), is shown by the solid line in Fig.~\ref{kapro4}.

\end{document}